\documentclass{svjour3}
\usepackage{makeidx,graphics,subfigure}
\usepackage{graphicx}

\begin{document}
\title{On the role of \emph{a priori} knowledge in the optimization of quantum information processing}

\author{Ming~Zhang \and
        Min~Lin \and
        S. G. Schirmer\and
        Hong-Yi Dai\and
        Zongtan~Zhou \and
         Dewen~Hu
}


\institute{Ming~Zhang, Min~Lin, Zongtan~Zhou,  Dewen~Hu \at
              College of Mechatronic Engineering
and Automation, National University of Defense Technology,
Changsha, Hunan 410073, People's Republic of China\\
              Tel.: 0086-731-84573335\\
              Fax: 0086-731-84573323\\
              \email{zhangming@nudt.edu.cn}           %  \\
                      \and
           S. G. Schirmer \at
            Department of Applied Mathematics and
Theoretical Physics, University of Cambridge, Wilberforce Road, CB3
0WA, UK\\
\and
        Hong-Yi Dai \at
        School of Science, National University of Defense Technology,
Changsha, Hunan 410073, People's Republic of China\\
}

\date{Received: date / Accepted: date}

\maketitle%

\begin{abstract}
%\boldmath
This paper explores   the role of \emph{a priori} knowledge in the
optimization of quantum information processing by investigating
optimum unambiguous discrimination problems for both the qubit and
qutrit states. In general, \emph{a priori} knowledge in optimum
unambiguous discrimination problems can be classed into two types:
\emph{a priori} knowledge of discriminated states themselves and
\emph{a priori} probabilities of preparing the states. It is
clarified that  whether \emph{a priori} probabilities of preparing
discriminated states are available or not, what type of
discriminators one should design just depends on what kind of the
classical knowledge of discriminated states. This is in contrast to
the observation that choosing the parameters of discriminators not
only relies on the \emph{a priori} knowledge of discriminated
states, but also depends on \emph{a priori} probabilities of
preparing the
 states. Two types of \emph{a priori} knowledge can
be utilized to improve optimum performance but  play the different
roles in the optimization from the view point of decision theory.
\end{abstract}

%\begin{IEEEkeywords}
%Decision theory, Quantum mechanics, Bayesian methods, a priori
%knowledge
%\end{IEEEkeywords}

% \fi

\section{Introduction}

Based on the observation that information is represented, stored,
processed, transmitted and readout by physical systems, Rolf
Landauer famously remarked that \emph{information is physical}.
Until recently, information was largely thought of in classical
terms with quantum mechanics playing at most a supportive role in
designing the equipment to store and process information.  However,
technological advances have ensured that quantum effects play an
increasingly important role.  This has motivated the birth of
quantum information theory~\cite{q1}, which is currently attracting
enormous interest due to its fundamental nature and potentially
important applications in quantum teleportation\cite{R1,R2}, dense
coding\cite{R4}, quantum cryptography\cite{R5,R6,R7}, and quantum
computation~\cite{c1,c2,c6}.

Quantum information theory can be considered as an extension and
generalization of classical information theory\cite{qc2} but there
are many important and special problems in this domain.  One of
these is the role of \emph{a priori} knowledge in quantum
information processing.
 Researchers have explored numerous ways to develop programmable
quantum devices~\cite{5} to accomplish various quantum information
processing tasks such as storing quantum dynamics in quantum
states\cite{8}, implementation of quantum maps\cite{9}, evaluating
the expectation value of any operator\cite{11} and quantum state
discrimination\cite{12,13,14,15,16}.  These methods are based on the
full utilization of \emph{a priori} knowledge. However, the role of
\emph{a priori} knowledge in quantum information processing has not
been fully explored so far, and this problem deserves the further
investigation.

How to make optimal decisions based on full utilization of \emph{a
priori} knowledge is an important issue in various domains,
especially in classical statistical decision theory~\cite{decision}.
To address the problem of the use of \emph{a priori} knowledge in
quantum information, we have to start with two basic questions: what
does \emph{a priori} knowledge mean in this domain and what role
does \emph{a priori} knowledge play in the optimization of quantum
information processing? Since quantum state
discrimination~\cite{4,A1,A2,A3} is very fundamental in the domain
of quantum information processing\cite{3}, it is reasonable to
carefully explore optimum unambiguous discrimination problems given
various types of \emph{a priori} knowledge.

In general, there are two types of \emph{a priori} knowledge in optimum
unambiguous discrimination problems: \emph{a priori} knowledge of
discriminated states themselves (\emph{a priori} knowledge of Type I)
and \emph{a priori} probabilities of preparing these states (\emph{a
priori} knowledge of Type II).  Type-I knowledge can be expressed in a
variety of forms.  Even when the discriminated states are classically
unknown, a single copy of them can be considered as a special form of
\emph{a priori} knowledge.  By making full use of a single copy of the
classically unknown states, Bergou and Hillery~\cite{12} constructed an
unambiguous discriminator, and Bergou et.al~\cite{16} showed how to
construct devices that can optimally discriminate between a classically
known and a classically unknown state.  Recently~\cite{qic}, the effect
of complete or incomplete \emph{a priori} classical knowledge of
discriminated states on the optimum unambiguous discrimination has been
investigated when \emph{a priori} probabilities of preparing the
discriminated states are known.

Under different conditions of \emph{a priori} knowledge, optimum
unambiguous discrimination problems are reduced to different
optimization problems.  What kind of optimization problems can optimum
unambiguous discrimination problems be reduced to?  This depends on
\emph{a priori} knowledge of both Type I and Type II. \emph{A priori}
probabilities of preparing discriminated states are presumed known in
many research papers~\cite{12,15,16} although optimal measurement for
quantum-state discrimination without a priori probabilities has also
been considered~\cite{ad}.  However, optimum unambiguous discrimination
problems have not been thoroughly investigated under various types of
\emph{a priori} knowledge so far.  Specially, the optimum unambiguous
discrimination problems have not been explored for multi-level quantum
states under different \emph{a priori} conditions.  Therefore, we would
like to exploit the role of \emph{a priori} knowledge in the
optimization by considering the optimum unambiguous discrimination
problems for both the qubit and qutrit states in this paper.  In this
paper, the effect of \emph{a priori} knowledge of both Type I and Type
II on the optimization will be examined carefully for both the qubit
states and qutrit states.  We will show how to design discriminators
that depend only on whether the classical knowledge of the discriminated
states is complete or incomplete. How to choose the parameters of the
discriminators, however, relies on \emph{a priori} knowledge of both
Type I and Type II.

The rest of this paper is organized as follows. In Sect. II, we
comprehensively present the results on optimal unambiguous
discrimination problems for two qubit states with and without different
\emph{a priori} preparing probability of the qubit states.  The
comparative analysis and some further discussions are present for
optimal unambiguous discrimination problems of two qubit states in
Sect. III. To further clarify the role of \emph{a priori} information in
the optimal decision and optimum success probability, we study the
optimal unambiguous discrimination for two qutrit states in Sect. IV.
The paper concludes with Sect. V.

\section{Optimal unambiguous discrimination problems for two qubit states}

To analysis the effect of \emph{a priori} information of the
discriminated states on optimum unambiguous discriminators
 and optimum success probabilities of   unambiguously discriminating two qubit states, we review the results on
 optimal
 unambiguous discrimination problems with    the knowledge of  \emph{a priori} preparing
 probabilities in the first three subsections. According to what kind of classical knowledge  can be
 utilized,  the 4 {cases}
 are discussed in the three subsections.

Case A1, without classical knowledge of either state but
 with a single copy of  unknown states;

Case A2,  with only classical knowledge of one of the two states and
a single copy of the other unknown state;

Case A3, with only classical knowledge of one of the two states and
the absolute value of the inner product of both states, and also
with a single copy of the other unknown state;

Case A4,  with classical knowledge of both states.

The A1 and A4 cases will be investigated in subsection A and C,
respectively, and the A2 and A3 cases will be studied in subsection
B.

  Since \emph{a priori} probabilities of preparing discriminated states are
often presumed known, optimal unambiguous discrimination problems
without \emph{a priori} preparing
 probability may be skipped to a certain degree, and have not been thoroughly investigated so far. In
the subsections D, E and F, we will further discuss various optimal
unambiguous discrimination problems without \emph{a priori}
preparing
 probability.
 We have also four cases taken into consideration as follow.

Case B1, without classical knowledge of either state but with
 a single copy of  unknown states;

Case B2, with classical knowledge of one of the two states and
 a single copy of the other unknown state;

Case B3, with classical knowledge of one of the two states and the
absolute value of the inner product of both states, and also with a
single copy of the other unknown state;

Case B4, with classical knowledge of both states.

The B1 and B4 cases will be investigated in subsection D and F,
respectively, and the B2 and B3 cases will be studied in subsection
E.
\subsection{Optimal unambiguous discrimination problems for Case A1}
In this subsection, we  review the result of Ref.  \cite{12}, and
further discuss the optimal unambiguous discrimination problems in
which the preparing probabilities is given, but none classical
knowledge of discriminated states is available.

 Given
two unknown quantum states $|\psi_{1}\rangle$ and
$|\psi_{2}\rangle$, we can construct a device to unambiguously
discriminate between them.
 Two classically
unknown states $|\psi_{1}\rangle$ and $|\psi_{2}\rangle$ are
provided as two inputs for two program registers, respectively. Then
we are given another qubit that is guaranteed to be one of two
unknown states $|\psi_{1}\rangle$ and $|\psi_{2}\rangle$ stored in
the two program registers. Our task is to determine, as best  we
can, which one the given qubit is. We are allowed to fail, but not
to make a mistake. What is the best procedure to accomplish this?
Our task is then reduced to the following measurement optimization
problem.

One has two input states
\begin{equation}
\label{A1-1}
|\Psi_{1}^{in}\rangle=|\psi_{1}\rangle_{A}|\psi_{1}\rangle_{B}|\psi_{2}\rangle_{C}
\end{equation}
and
\begin{equation}
\label{A1-2}
|\Psi_{2}^{in}\rangle=|\psi_{1}\rangle_{A}|\psi_{2}\rangle_{B}|\psi_{2}\rangle_{C}
\end{equation}
 where the subscripts $A$ and $C$ refer to the program registers, and the subscript
$B$ refers to the data register. Our goal is to unambiguously
distinguish between these inputs.

Let the elements of our POVM (positive-operator-valued measure) be
$\Pi_{1}$, corresponding to unambiguously detecting
$|\psi_{1}\rangle$,  $\Pi_{2}$,
 corresponding to unambiguously detecting
 $|\psi_{2}\rangle$,  and
$\Pi_{0}$,  corresponding to  failure, respectively. The
probabilities of successfully identifying the two possible input
states are given by
\begin{equation}
\label{A2-1}
\langle\Psi_{1}^{in}|{\Pi}_{1}|\Psi_{1}^{in}\rangle=p_{1}
\end{equation}
and
\begin{equation}
\label{A2-2}
\langle\Psi_{2}^{in}|{\Pi}_{2}|\Psi_{2}^{in}\rangle=p_{2}
\end{equation}
and the condition of no errors implies that
\begin{equation}
\label{A3} {\Pi}_{1}|\Psi_{2}^{in}\rangle=0;\
{\Pi}_{2}|\Psi_{1}^{in}\rangle=0
\end{equation}
In addition, because the alternatives represented by the POVM
exhaust all possibilities, we have that
\begin{equation}
\label{A4} {\Pi}_{1}+ {\Pi}_{2}+ {\Pi}_{0}=I
\end{equation}
 Since we  have no classical knowledge
 about $|\psi_{1}\rangle$ and $|\psi_{2}\rangle$,
 the right way of constructing POVM operators is  to  take advantage of the
 symmetrical
properties of the states. Denoting $|0\rangle$ and $|1\rangle$ as
two vectors of a basis, we define the antisymmetric state
\begin{equation}
\label{A5-1}
|\psi_{BC}^{-}\rangle=\frac{1}{\sqrt{2}}(|0\rangle_{B}|1\rangle_{C}-|1\rangle_{B}|0\rangle_{C})
\end{equation}
and
\begin{equation}
\label{A5-2}
|\psi_{AB}^{-}\rangle=\frac{1}{\sqrt{2}}(|0\rangle_{A}|1\rangle_{B}-|1\rangle_{A}|0\rangle_{B})
\end{equation}
and introduce the projectors to the antisymmetric subspaces of the
corresponding qubit as
\begin{equation}
\label{A6-1}  P_{BC}^{as}=|\psi_{BC}^{-}\rangle\langle\psi_{BC}^{-}|
\end{equation}
and
\begin{equation}
\label{A6-2}
{P}_{AB}^{as}=|\psi_{AB}^{-}\rangle\langle\psi_{AB}^{-}|
\end{equation}
We now can take for ${\Pi}_{1}$ and ${\Pi}_{2}$ operators
\begin{equation}
\label{A7-1} {\Pi}_{1}=\lambda_{1}I_{A}\otimes{P}_{BC}^{as}
\end{equation}
and
\begin{equation}
\label{A7-2} {\Pi}_{2}=\lambda_{2}{P}_{AB}^{as}\otimes{I_{C}}
\end{equation}
To assure that ${\Pi}_{1}$, ${\Pi}_{2}$ and
${\Pi}_{0}=I-{\Pi}_{1}-{\Pi}_{2}$ be semi-positive operators, the
following constraints should be satisfied:
\begin{equation}
\label{A7} 2-\lambda_{1}-\lambda_{2}\geq0;\
1-\lambda_{1}-\lambda_{2}+\frac{3}{4}\lambda_{1}\lambda_{2}\geq0
\end{equation}

 After some calculations, we have
$p_{i}=\frac{1}{2}\lambda_{i}(1-{\beta}^{2})$, where $i=1,2$ and
$\beta=|\langle\psi_{1}|\psi_{2}\rangle|$. Suppose that $\eta_{1}$
is the preparation probability of $|\psi_{1}\rangle$, the average
success probability is $P=p_{1}\eta_{1}+p_{2}(1-\eta_{1})$.

 Since
we have knowledge of $\eta_{1}$, our task is reduced to designing
$\lambda_{1}(\eta_{1})$ and $\lambda_{2}(\eta_{1})$ such that the
following average success probability
\begin{equation}
\label{A8}
P={\frac{1}{2}[\lambda_{1}\eta_{1}+\lambda_{2}(1-\eta_{1})](1-{\beta}^{2})}
\end{equation}
is maximal with the constrains given by Eq. (\ref{A7}). This is to
say, the loss function can be expressed as
\begin{equation}
\label{A8-J}
J=\max\min_{\{\eta_{1}\}}\{{\frac{1}{2}[\lambda_{1}\eta_{1}+\lambda_{2}(1-\eta_{1})](1-{\beta}^{2})}\}
\end{equation}

  In this case, the
optimum success probability has been summarized as follows
\begin{equation}
\label{A-9} P_{0}^{opt}(\beta,\eta_{1})=\ \bigg\{\begin{array}{cc}
\frac{1}{2}(1-\eta_{1})(1-{\beta}^{2})&\eta_{1}\leq{\frac{1}{5}}\\
\frac{2}{3}[1-\sqrt{\eta_{1}(1-\eta_{1})}](1-{\beta}^{2})&\frac{1}{5}\leq{\eta_{1}}\leq{\frac{4}{5}}\\
\frac{1}{2}\eta_{1}(1-{\beta}^{2})&\eta_{1}\geq{\frac{4}{5}}
\end{array}
\end{equation}
where the subscript $0$ of $P_{0}^{opt}$ means that we have no
\emph{a priori} classical knowledge of $|\psi_{1}\rangle$ and
$|\psi_{2}\rangle$ and the corresponding optimal action parameters
are given by
\begin{equation}
\label{A-9a} \lambda_{1}^{0,opt}(\eta_{1})=\
\bigg\{\begin{array}{cc}
0&\eta_{1}\leq{\frac{1}{5}}\\
\frac{2}{3}[2-\sqrt{\frac{1-\eta_{1}}{\eta_{1}}}]&\frac{1}{5}\leq{\eta_{1}}\leq{\frac{4}{5}}\\
1&\eta_{1}\geq{\frac{4}{5}}
\end{array}
\end{equation}
and
\begin{equation}
\label{A-9b} \lambda_{2}^{0,opt}(\eta_{1})=\
\bigg\{\begin{array}{cc}
1&\eta_{1}\leq{\frac{1}{5}}\\
\frac{2}{3}[2-\sqrt{\frac{\eta_{1}}{1-\eta_{1}}}]&\frac{1}{5}\leq{\eta_{1}}\leq{\frac{4}{5}}\\
0&\eta_{1}\geq{\frac{4}{5}}
\end{array}
\end{equation}

\textbf{Remark:}  We would like to underline that the we make the
optimal decision in Eqs. (\ref{A-9a}) and (\ref{A-9b}) without the
knowledge $\beta$, but the optimum success probability (\ref{A-9})
is the function of both $\beta$ and $\eta_{1}$.

\subsection{Optimal unambiguous discrimination problems for Cases A2 and A3}
In this subsection, we re-discuss the optimal unambiguous
discrimination problem for the   {cases} A2 and A3 from the view
point of decision theory, and throw some new insights, which are
different from those in Ref. \cite{16,qic}.

Given one known quantum state $|\psi_{1}\rangle$ and one unknown
quantum state $|\psi_{2}\rangle$, we can construct a device that
 unambiguously discriminate between them.
 We shall consider the following problem which may be a simple
version of a programmable state discriminator. The unknown state
$|\psi_{2}\rangle$ is provided as an input  for the program
register.  Then we are given another qubit that is guaranteed to be
in  the known state $|\psi_{1}\rangle$  or the unknown state
$|\psi_{2}\rangle$ stored in the program register. Our task is to
determine, as best we can, which one the given qubit is. As in case
A1, we are allowed to fail, but not to make a mistake. What is the
best procedure to accomplish this?

In line with Ref. \cite{12},  one can construct such a device by
viewing this problem as a task in measurement optimization. The
measurement is allowed to return an inconclusive result but never an
erroneous one. Thus, it will be described by a POVM
 that will return outcome $1$ (the
unknown state stored in the data register matches the known state
$|\psi_{1}\rangle$),  $2$ (the unknown state stored in the data
register matches $|\psi_{2}\rangle$  in the program register), or
$0$ (we do not learn anything about the unknown state stored in the
data register). Our task is then reduced to the following
measurement optimization problem.

One has two input states
\begin{equation}
\label{1a}
|\Psi_{1}^{in}\rangle=|\psi_{2}\rangle_{A}|\psi_{1}\rangle_{B}
\end{equation}
and
\begin{equation}
\label{1b}
|\Psi_{2}^{in}\rangle=|\psi_{2}\rangle_{A}|\psi_{2}\rangle_{B}
\end{equation}
 where the subscript $A$  refers to the program register  ($A$ contains $|\psi_{2}\rangle$), and the subscript
$B$ refers to the data register. Our goal is to unambiguously
distinguish between these inputs.

Let the elements of our POVM be $\Pi_{1}$, corresponding to
unambiguously detecting $|\psi_{1}\rangle$,  $\Pi_{2}$,
 corresponding to unambiguously detecting
 $|\psi_{2}\rangle$,  and
$\Pi_{0}$,  corresponding to  failure, respectively. The
probabilities of successfully identifying the two possible input
states are given by
\begin{equation}
\label{2a} \langle\Psi_{1}^{in}|{\Pi}_{1}|\Psi_{1}^{in}\rangle=p_{1}
\end{equation}
and
\begin{equation} \label{2b}
\langle\Psi_{2}^{in}|{\Pi}_{2}|\Psi_{2}^{in}\rangle=p_{2}
\end{equation}
and the condition of no errors implies that
\begin{equation}
\label{3} {\Pi}_{1}|\Psi_{2}^{in}\rangle=0;\
{\Pi}_{2}|\Psi_{1}^{in}\rangle=0
\end{equation}
In addition, because the alternatives represented by the POVM
exhaust all possibilities, we have that
\begin{equation}
\label{4} {\Pi}_{1}+ {\Pi}_{2}+ {\Pi}_{0}=I
\end{equation}
 Since we  know nothing
 about $|\psi_{2}\rangle$ but have the classical knowledge of $|\psi_{1}\rangle$,
 the right way of constructing POVM operators is  to  take advantage of the
 symmetrical
properties of the state as well as the classical knowledge of
$|\psi_{1}\rangle$. Denoting $|\psi_{1}^{\perp}\rangle$ as the unit
vector orthogonal to $|\psi_{1}\rangle$, we  define the
antisymmetric state
\begin{equation}
\label{5}
|\psi_{AB}^{-}\rangle=\frac{1}{\sqrt{2}}(|\psi_{1}\rangle_{A}|\psi_{1}^{\perp}\rangle_{B}-|\psi_{1}^{\perp}\rangle_{A}|\psi_{1}\rangle_{B})
\end{equation}
and introduce the projectors to the antisymmetric subspaces of the
corresponding qubit as
\begin{equation}
\label{6} {P}_{AB}^{as}=|\psi_{AB}^{-}\rangle\langle\psi_{AB}^{-}|
\end{equation}

\textbf{Remark:} suppose $|\psi_{1}\rangle$ can be expressed in
terms of a basis of $|0\rangle$ and $|1\rangle$ as
$|\psi_{1}\rangle=\cos\frac{\theta}{2}|0\rangle+e^{i\phi}\sin\frac{\theta}{2}|1\rangle$
where   $0\leq\phi<2\pi$   and $0\leq\theta\leq\pi$, then we can
choose
$|\psi_{1}^{\perp}\rangle=\sin\frac{\theta}{2}|0\rangle-e^{i\phi}\cos\frac{\theta}{2}|1\rangle$.

By making full use of  the knowledge of $|\psi_{1}\rangle$ and
$|\psi_{1}^{\perp}\rangle$,   we  construct the measurement
operators ${\Pi}_{1}$ and ${\Pi}_{2}$ to satisfy the no-error
condition given by Eq.(\ref{3}) as follows:
\begin{equation}
\label{7-1} {\Pi}_{1}=\lambda_{1}{P}_{AB}^{as}
\end{equation}
and
\begin{equation}
\label{7-2}
{\Pi}_{2}=\lambda_{2}{|\psi_{1}\rangle_{A}|\psi_{1}^{\perp}\rangle_{B}{_{B}\langle}\psi_{1}^{\perp}|_{A}\langle\psi_{1}|}+\lambda_{3}{|\psi_{1}^{\perp}\rangle_{A}|\psi_{1}^{\perp}\rangle_{B}{_{B}\langle}\psi_{1}^{\perp}|_{A}\langle\psi_{1}^{\perp}|}
\end{equation}
where  $\lambda_{1}$, $\lambda_{2}$ and $\lambda_{3}$ are
undetermined nonnegative real numbers. Using the  Eqs. (\ref{7-1})
and (\ref{7-2}), we have
\begin{equation}
\label{8}
p_{1}=\langle\Psi_{1}^{in}|{\Pi}_{1}|\Psi_{1}^{in}\rangle=\frac{1}{2}\lambda_{1}(1-\beta^{2})
\end{equation}
\begin{equation}
\label{9}
p_{2}=\langle\Psi_{2}^{in}|{\Pi}_{2}|\Psi_{2}^{in}\rangle=\lambda_{2}\beta^{2}(1-\beta^{2})+\lambda_{3}(1-\beta^{2})^{2}
\end{equation}

By assuming that the  preparation probabilities of
$|\psi_{1}\rangle$ and $|\psi_{2}\rangle$ are $\eta_{1}$ and
$\eta_{2}$ (where $\eta_{2}=1-\eta_{1}$), respectively, we can
define the average probability $P$ of successfully discriminating
two states as
\begin{equation}
\label{10}
P=[\frac{1}{2}\lambda_{1}\eta_{1}+\lambda_{2}\beta^{2}\eta_{2}+\lambda_{3}(1-\beta^{2})\eta_{2}](1-\beta^{2})
\end{equation}
where $\beta=|\langle\psi_{1}|\psi_{2}\rangle|$, and our task is to
maximize the performance Eq. (\ref{10}) subject to the constraint
that ${\Pi}_{0}=I-{\Pi}_{1}-{\Pi}_{2}$ is a positive operator.

To assure that ${\Pi}_{0}$, ${\Pi}_{1}$ and ${\Pi}_{2}$ are positive
operators, we have the following inequality constraints:

\begin{equation}
\label{i1}
1-{\lambda_{1}}-{\lambda_{2}}+\frac{1}{2}{\lambda_{1}}{\lambda_{2}}\geq0
\end{equation}
and
\begin{equation}
\label{i2} 0\leq{\lambda_{i}}\leq1(i=1,2,3)
\end{equation}

Subsequently, we will discuss our strategies  for the A2 and A3
cases.

\textbf{(i)} For the A2 case,  we have the knowledge of preparing
probability $\eta_{1}$, but no knowledge of $\beta$. It can be
demonstrated that one cannot maximize Eq. (\ref{10}) everywhere
simultaneously without the classical knowledge of $\beta$. Still, we
can give some further analysis.

Our strategy  is to design $\lambda^{1,w\beta}_{1}(\eta)$,
$\lambda^{1,w\beta}_{2}(\eta)$ and $\lambda^{1,w\beta}_{3}(\eta)$ to
optimize the performance
\begin{equation}
\label{H10-1}
J=\max\min_{\{\beta\}}[\frac{1}{2}\lambda_{1}\eta_{1}+\lambda_{2}(1-\eta_{1})+(\lambda_{3}-\lambda_{2})(1-\beta^{2})(1-\eta_{1})]
\end{equation}
subject to the  constraints described by Eqs. (\ref{i1})  and
(\ref{i2}).

No matter what $\eta_{1}$ is, one should  always choose
$\lambda^{1,w\beta}_{3}(\eta_{1})=1$. As for
$\lambda^{1,w\beta}_{1}(\eta_{1})$ and
$\lambda^{1,w\beta}_{2}(\eta_{1})$, the problem is reduced to
 maximizing
$\frac{1}{2}\lambda_{1}\eta_{1}+\lambda_{2}(1-\eta_{1})$ subject to
the  constraints described by Eqs. (\ref{i1})  and (\ref{i2}).

After some calculations, we have
\begin{equation}
\label{24-a} \lambda_{1}^{1,w\beta}(\eta_{1})=\
\bigg\{\begin{array}{cc}
0&\eta_{1}\leq{\frac{1}{2}}\\
2(1-\sqrt{\frac{1-\eta_{1}}{\eta_{1}}})&\frac{1}{2}\leq{\eta_{1}}\leq{\frac{4}{5}}\\
1&\eta_{1}\geq{\frac{4}{5}}
\end{array}
\end{equation}
and
\begin{equation}
\label{24-b} \lambda_{2}^{1,w\beta}(\eta_{1})=\
\bigg\{\begin{array}{cc}
1&\eta_{1}\leq{\frac{1}{2}}\\
2-\sqrt{\frac{\eta_{1}}{1-\eta_{1}}}&\frac{1}{2}\leq{\eta_{1}}\leq{\frac{4}{5}}\\
0&\eta_{1}\geq{\frac{4}{5}}
\end{array}
\end{equation}
and
\begin{equation}
\label{24-c} \lambda^{1,w\beta}_{3}(\eta_{1})=1
\end{equation}

 By substituting
$\lambda^{1,w\beta}_{1}(\eta_{1})$,
$\lambda^{{1,w\beta}}_{2}(\eta_{1})$ and
$\lambda^{1,{w\beta}}_{3}(\eta_{1})$ into (\ref{10}), we obtain the
actual optimum success probability in this strategy:
\begin{equation}
\label{24-1} P_{1}^{w\beta}(\beta,\eta_{1})=\
\bigg\{\begin{array}{cc}
P_{1_1}^{w\beta}(\beta,\eta_{1})&\eta_{1}\leq{\frac{1}{2}}\\
P_{1_2}^{w\beta}(\beta,\eta_{1})&\frac{1}{2}\leq{\eta_{1}}\leq{\frac{4}{5}}\\
P_{1_3}^{w\beta}(\beta,\eta_{1})&\eta_{1}\geq{\frac{4}{5}}
\end{array}
\end{equation}
with
\begin{equation}
\label{24-1a}
P_{1_1}^{w\beta}(\beta,\eta_{1})=(1-\eta_{1})(1-{\beta}^{2})
\end{equation}
and
\begin{equation}
\label{24-1b}
P_{1_2}^{w\beta}(\beta,\eta_{1})=[1+{\beta}^{2}(1-\eta_{1})-(1+{\beta}^{2})\sqrt{\eta_{1}(1-\eta_{1})}](1-{\beta}^{2})
\end{equation}
and
\begin{equation}
\label{24-1c} P_{1_3}^{w\beta}(\beta,\eta_{1})=
[1-\frac{1}{2}\eta_{1}-{\beta}^{2}(1-\eta_{1})](1-{\beta}^{2})
\end{equation}
 where  the subscript $1$ of $P_{1}^{w\beta}$ means that we just
have \emph{a priori} classical knowledge of $|\psi_{1}\rangle$, one
of two discriminated states, and the superscript $w\beta$ of
$P_{1}^{w\beta}$ implies that the  actual optimum success
probability is obtain for the worst case  of $\beta$.

\textbf{Remark:} In this case, our ``optimal" actions or decisions
in Eqs. (\ref{24-a}-\ref{24-c}) can be only considered as the
function of  the preparing probability $\eta_{1}$ but the actual
success probability given by Eq. (\ref{24-1}) with Eqs.
(\ref{24-1a}-\ref{24-1c}) is still the function of both $\beta$ and
$\eta_{1}$ even when we have no idea about $\beta$.

\textbf{(ii) }With \emph{a priori} classical knowledge of both
$|\langle\psi_{1}|\psi_{2}\rangle|=\beta$  and $\eta_{1}$ in hand,
our task in the third case is to get the optimum values
$\lambda^{1+,opt}_{1}(\beta,\eta_{1})$,
$\lambda^{1+,opt}_{2}(\beta,\eta_{1})$ and
$\lambda^{1+,opt}_{3}(\beta,\eta_{1})$ to optimize the  average
success probability
\begin{equation}
\label{A10}
J=[\frac{1}{2}\lambda_{1}\eta_{1}+\lambda_{2}\beta^{2}(1-\eta_{1})+\lambda_{3}(1-\beta^{2})(1-\eta_{1})](1-\beta^{2})
\end{equation}
 subject to the constraints Eqs. (\ref{i1}) and (\ref{i2}).

After some calculations,  we have
\begin{equation}
\label{A28a} \lambda^{1+,opt}_{1}(\beta,\eta_{1})=\
\bigg\{\begin{array}{cc}
0&\eta_{1}\leq\frac{{\beta}^{2}}{1+{\beta}^{2}}\\
2(1-\beta\sqrt{\frac{1-\eta_{1}}{\eta_{1}}})&\frac{{\beta}^{2}}{1+{\beta}^{2}}\leq{\eta_{1}}\leq{\frac{4{\beta}^{2}}{1+4{\beta}^{2}}}\\
1&\eta_{1}\geq{{\frac{4{\beta}^{2}}{1+4{\beta}^{2}}}}
\end{array}
\end{equation}
and
\begin{equation}
\label{A28b} \lambda^{1+,opt}_{2}(\beta,\eta_{1})=\
\bigg\{\begin{array}{cc}
1&\eta_{1}\leq\frac{{\beta}^{2}}{1+{\beta}^{2}}\\
2-\frac{1}{\beta}\sqrt{\frac{\eta_{1}}{1-\eta_{1}}}&\frac{{\beta}^{2}}{1+{\beta}^{2}}\leq{\eta_{1}}\leq{\frac{4{\beta}^{2}}{1+4{\beta}^{2}}}\\
0&\eta_{1}\geq{{\frac{4{\beta}^{2}}{1+4{\beta}^{2}}}}
\end{array}
\end{equation}
and
\begin{equation}
\label{A28c} \lambda^{1+,opt}_{3}(\beta,\eta_{1})\equiv1
\end{equation}

Taking Eq. (\ref{A10}) into consideration, we obtain the
corresponding optimum success probabilities:
\begin{equation}
\label{A28-1} P_{1+}^{opt}(\beta,\eta_{1})=\
\bigg\{\begin{array}{cc}
P_{1+_1}^{opt}(\beta,\eta_{1})&\eta_{1}\leq\frac{{\beta}^{2}}{1+{\beta}^{2}}\\
P_{1+_2}^{opt}(\beta,\eta_{1})&\frac{{\beta}^{2}}{1+{\beta}^{2}}\leq{\eta_{1}}\leq{\frac{4{\beta}^{2}}{1+4{\beta}^{2}}}\\
P_{1+_3}^{opt}(\beta,\eta_{1})&\eta_{1}\geq{{\frac{4{\beta}^{2}}{1+4{\beta}^{2}}}}
\end{array}
\end{equation}
with
\begin{equation}
\label{A28-1a} P_{1+_1}^{opt}(\beta,\eta_{1})=
(1-\eta_{1})(1-\beta^{2})
\end{equation}
and
\begin{equation}
\label{A28-1b}
P_{1+_2}^{opt}(\beta,\eta_{1})=[1+{\beta}^{2}(1-\eta_{1})-2\beta\sqrt{\eta_{1}(1-\eta_{1})}](1-\beta^{2})
\end{equation}
and
\begin{equation}
\label{A28-1c}
P_{1+_3}^{opt}(\beta,\eta_{1})=[1-\frac{1}{2}\eta_{1}-\beta^{2}(1-\eta_{1})](1-\beta^{2})
\end{equation}
 where the subscript $1+$ of $P_{1+}^{opt}$ means
that we  have \emph{a priori} classical knowledge of   one of the
two discriminated states and the absolute value of the inner product
of the two states.

\textbf{Remark:} In the Case A3, both the optimal actions or
decisions in Eqs. (\ref{A28a}-\ref{A28c}) and the actual success
probability given by Eq. (\ref{A28-1}) with Eqs.
(\ref{A28-1a}-\ref{A28-1c}) can be considered as the functions of
both $\beta$ and $\eta_{1}$.

\subsection{Optimal unambiguous discrimination problems for A4}

In this subsection, we recall the result of Ref. \cite{3,A4} for the
A4 case.

If we have complete \emph{a priori} classical knowledge of both
$|\psi_{1}\rangle$ and $|\psi_{2}\rangle$, the measurement is
performed on  the detected qubit.  One can select the detection
operators as
\begin{equation}
\label{c1a}
{\Pi}_{1}=\lambda_{1}{|\psi_{2}^{\perp}\rangle\langle\psi_{2}^{\perp}|}
\end{equation}
and
\begin{equation}
\label{c1b}
{\Pi}_{2}=\lambda_{2}{|\psi_{1}^{\perp}\rangle\langle\psi_{1}^{\perp}|}
\end{equation}

Our task is to choose  $\lambda_{1}$ and $\lambda_{2}$  based on
\emph{a priori} information such that the average success
probability
\begin{equation}
\label{c2}
P=[\lambda_{1}\eta_{1}+\lambda_{2}(1-\eta_{1})](1-\beta^{2})
\end{equation}
is maximized. This is to say, the loss function can be expressed as
\begin{equation}
\label{c2-1}
J=\max\min_{\{\eta_{1}\}}\{{\frac{1}{2}[\lambda_{1}\eta_{1}+\lambda_{2}(1-\eta_{1})](1-{\beta}^{2})}\}
\end{equation}

To assure that ${\Pi}_{0}$, ${\Pi}_{1}$ and ${\Pi}_{2}$ are positive
operators, we have the following inequality constraints:
\begin{equation}
\label{c3} 1-\lambda_{1}-\lambda_{2}\beta^{2}\geq0
\end{equation}
and
\begin{equation} \label{c4}
1-\lambda_{1}-\lambda_{2}+(1-\beta^{2})\lambda_{1}\lambda_{2}\geq0
\end{equation}
where $|\langle\psi_{1}|\psi_{2}\rangle|=\beta$.

Since we have knowledge of preparing probability $\eta_{1}$ and
$\beta$, we will make the following decision
\begin{equation}
\label{28a} \lambda_{1}^{2,opt}(\beta,\eta_{1})=\
\bigg\{\begin{array}{cc}
0&\eta_{1}\leq{\frac{\beta^{2}}{1+\beta^{2}}}\\
\frac{1}{1-\beta^{2}}(1-\beta\sqrt{\frac{1-\eta_{1}}{\eta_{1}}})&\frac{\beta^{2}}{1+\beta^{2}}\leq{\eta_{1}}\leq{\frac{1}{1+\beta^{2}}}\\
1&\eta_{1}\geq{\frac{1}{1+\beta^{2}}}
\end{array}
\end{equation}
and
\begin{equation}
\label{28b} \lambda_{2}^{2,opt}(\beta,\eta_{1})=\
\bigg\{\begin{array}{cc}
1&\eta_{1}\leq{\frac{\beta^{2}}{1+\beta^{2}}}\\
\frac{1}{1-\beta^{2}}(1-\beta\sqrt{\frac{\eta_{1}}{1-\eta_{1}}})&\frac{\beta^{2}}{1+\beta^{2}}\leq{\eta_{1}}\leq{\frac{1}{1+\beta^{2}}}\\
0&\eta_{1}\geq{\frac{1}{1+\beta^{2}}}
\end{array}
\end{equation}
 Furthermore, we can obtain the optimum success probability for this case (also as
per Ref. \cite{3,A4}).
\begin{equation}
\label{28-1} P_{2}^{opt}(\beta,\eta_{1})=\
\bigg\{\begin{array}{cc} P_{2_1}^{opt}(\beta,\eta_{1})&\eta_{1}\leq{\frac{\beta^{2}}{1+\beta^{2}}}\\
P_{2_2}^{opt}(\beta,\eta_{1})&\frac{\beta^{2}}{1+\beta^{2}}\leq{\eta_{1}}\leq{\frac{1}{1+\beta^{2}}}\\
P_{2_3}^{opt}(\beta,\eta_{1})&\eta_{1}\geq{\frac{1}{1+\beta^{2}}}
\end{array}
\end{equation}
with
\begin{equation} \label{28-1a} P_{2_1}^{opt}(\beta,\eta_{1})=(1-\eta_{1})(1-\beta^{2})
\end{equation}
and
\begin{equation}
\label{28-1b} P_{2_2}^{opt}(\beta,\eta_{1})=
1-2\sqrt{\eta_{1}(1-\eta_{1})}\beta
\end{equation}
and
\begin{equation}
\label{28-1c} P_{2_3}^{opt}(\beta,\eta_{1})= \eta_{1}(1-\beta^{2})
\end{equation}
where $|\langle\psi_{1}|\psi_{2}\rangle|=\beta$, the subscript $2$
of $P_{2}^{opt}$ means that we  have the classical knowledge of both
discriminated states.

\textbf{Remark:} For the Case A4, it is not surprising to find that
the optimal actions or decisions in Eqs. (\ref{28a}-\ref{28b}) and
the optimal success probability given by Eq. (\ref{28-1}) with Eqs.
(\ref{28-1a}-\ref{28-1c}) are the function of both $\beta$ and
$\eta_{1}$.

\subsection{Optimal unambiguous discrimination problems for Case B1}
Since we have the same classical knowledge of discriminated states
in this case as in Section II. A,  we can follow the analysis in
Section II. A and choose ${\Pi}_{1}$ and ${\Pi}_{2}$ as Eqs.
(\ref{A7-1}) and (\ref{A7-2}).

To assure that ${\Pi}_{1}$, ${\Pi}_{2}$ and
${\Pi}_{0}=I-{\Pi}_{1}-{\Pi}_{2}$ be semi-positive operators, the
 constraints   on $\lambda_{1}$ and $\lambda_{2}$ described by Eq.(\ref{A7})
 should be satisfied.

However, since we have no knowledge of preparing probability, we
have to design $\lambda_{1}$ and $\lambda_{2}$ without \emph{a
priori} information of $\eta_{1}$. Our strategy is to maximize   the
minimal performance
\begin{equation}
\label{A-10}
J=P_{0}^{w\eta_{1}}(\beta)=\max\min_{\{\eta_{1}\}}{\frac{1}{2}}[\lambda_{1}\eta_{1}+\lambda_{2}(1-\eta_{1})](1-{\beta}^{2})
\end{equation}
with the constraints in Eq. (\ref{A7}).

After careful calculations, we obtain that
\begin{equation}
\label{A-10a}
\lambda^{{0,w\eta_{1}}}_{1}=\lambda^{0,w\eta_{1}}_{2}=\frac{2}{3}
\end{equation}
Substituting Eq. (\ref{A-10a}) into Eq. (\ref{A-10}) yields
\begin{equation}
\label{A-11} P_{0}^{w\eta_{1}}(\beta)=\frac{1}{3}(1-{\beta}^{2})
\end{equation}

\textbf{Remark:} It should be pointed out that the optimal action or
decision in Eq. (\ref{A-10a}) is constant and the optimum success
probability is the function of $\beta$ in Case B1.

\subsection{Optimal unambiguous discrimination problems for Cases B2 and B3}
In this subsection, we will discuss the optimal unambiguous
discrimination problems for the B2  and B3 {cases} where partial
classical knowledge   but none  knowledge of preparing probabilities
of discriminated states are available.

Since we have the same partial classical knowledge of discriminated
states in this section as in Section II.B,  we can follow the
analysis in Section II.B and choose ${\Pi}_{1}$ and ${\Pi}_{2}$ as
Eqs.(\ref{7-1}) and (\ref{7-2}).

To assure that ${\Pi}_{1}$, ${\Pi}_{2}$ and
${\Pi}_{0}=I-{\Pi}_{1}-{\Pi}_{2}$ be semi-positive operators, the
 constraints on $\lambda_{1}$, $\lambda_{2}$ and $\lambda_{3}$ described by (\ref{i1}) and  (\ref{i2}) should be satisfied.

Our task is to design $\lambda_{1}$, $\lambda_{2}$ and $\lambda_{3}$
such that the average success probability given by Eq. (\ref{10}) is
maximized.

Subsequently, we will discuss our strategies for the B2 and B3
cases, respectively.

(i) If we  have neither the knowledge of preparing probabilities nor
the knowledge of $\beta$, our task is reduced to designing
$\lambda_{1}^{1,w\beta\eta_{1}}$, $\lambda_{2}^{1,w\beta\eta_{1}}$
and $\lambda_{3}^{1,w\beta\eta_{1}}$ to optimize the performance
\begin{equation}
\label{BH10-1}
J=\max\min_{\{\beta,\eta_{1}\}}[\frac{1}{2}\lambda_{1}\eta_{1}+\lambda_{2}(1-\eta_{1})+(\lambda_{3}-\lambda_{2})(1-\beta^{2})(1-\eta_{1})]
\end{equation}
subject to the  constraints in Eqs. (\ref{i1})  and (\ref{i2}).

Following some similar calculations in the subsection II. B, we have
the optimal actions as follows
\begin{equation}
\label{BH10a} \lambda_{1}^{1,w\beta\eta_{1}}=3-\sqrt{5}
\end{equation}
and
\begin{equation}
\label{BH10b} \lambda_{2}^{1,w\beta\eta_{1}}=\frac{1}{2}(3-\sqrt{5})
\end{equation}
and
\begin{equation}
\label{BH10c} \lambda_{3}^{1,w\beta\eta_{1}}=1
\end{equation}

 By substituting them  into Eq. (\ref{10}),
we get the  actual  success probability with regard to  this
strategy:
\begin{equation}
\label{W24-1}
P_{1}^{w\beta\eta_{1}}(\beta,\eta_{1})=[\frac{3-\sqrt{5}}{2}+\frac{\sqrt{5}-1}{2}(1-{\beta}^{2})(1-\eta_{1})](1-{\beta}^{2})
\end{equation}
where  the subscript $1$ of $P_{1}^{w\beta\eta_{1}}$ means that we
just have \emph{a priori} classical knowledge of $|\psi_{1}\rangle$,
one of two discriminated states, and the superscript
$w\beta\eta_{1}$ implies that the  actual ``optimum" success
probability is  based on the choice of the parameters of measurement
operators for the worst case of both $\beta$ and $\eta_{1}$.

\textbf{Remark: }It is interesting to underline that the optimal
decision or action is independent of both $\beta$ and $\eta_{1}$ but
the actual ``optimum" success probability based on the decision is
still the function of both $\beta$ and $\eta_{1}$.

(ii) For the B3 case,  we have the knowledge of $\beta$, but no
knowledge of preparing probability $\eta_{1}$.

Our task is to design $\lambda^{1+,w\eta_{1}}_{1}(\beta)$,
$\lambda^{1+,w\eta_{1}}_{2}(\beta)$ and
$\lambda^{1+,w\eta_{1}}_{3}(\beta)$ to maximize the minimal
performance
\begin{equation}
\label{10-1}
J=\max\min_{\{\eta_{1}\}}[\frac{1}{2}\lambda_{1}\eta_{1}+\lambda_{2}(1-\eta_{1})+(\lambda_{3}-\lambda_{2})(1-\beta^{2})(1-\eta_{1})]
\end{equation}
subject to the  constraints given by Eqs. (\ref{i1})  and
(\ref{i2}).

After some calculations, we have
\begin{equation}
\label{wA28a} \lambda_{1}^{1+,w\eta_{1}}(\beta)=\
\bigg\{\begin{array}{cc}
1&\beta\leq\frac{\sqrt{2}}{2}\\
\beta^{2}+2-\sqrt{\beta^{4}+4\beta^{2}}&\beta\geq\frac{\sqrt{2}}{2}
\end{array}
\end{equation}
and
\begin{equation}
\label{wA28b} \lambda_{2}^{1+,w\eta_{1}}(\beta)=\
\bigg\{\begin{array}{cc}
0&\beta\leq\frac{\sqrt{2}}{2}\\
\frac{3}{2}-\sqrt{\frac{1}{4}+\frac{1}{\beta^{2}}}&\beta\geq\frac{\sqrt{2}}{2}
\end{array}
\end{equation}
and
\begin{equation}
\label{wA28c} \lambda_{3}^{1+,w\eta_{1}}(\beta)\equiv1
\end{equation}

By substituting $\lambda^{1+,w\eta_{1}}_{1}$,
$\lambda^{{1+,w\eta_{1}}}_{2}$ and $\lambda^{1+,{w\eta_{1}}}_{3}$
into (\ref{10}), we obtain the actual success probability:
\begin{equation}
\label{wA28-1} P_{1+}^{A}(\beta,\eta_{1})=\ \bigg\{\begin{array}{cc}
[\frac{1}{2}+(\frac{1}{2}-{\beta}^{2})(1-\eta_{1})](1-{\beta}^{2})&\beta\leq\frac{\sqrt{2}}{2}\\
(1+\frac{1}{2}{\beta}^{2}-\frac{1}{2}\sqrt{{\beta}^{4}+4{\beta}^{2}})(1-{\beta}^{2})&\beta\geq\frac{\sqrt{2}}{2}
\end{array}
\end{equation}
and optimum success probabilities for  the worst case
\begin{equation}
\label{wA28-2} P_{1+}^{w\eta_{1}}(\beta)=\ \bigg\{\begin{array}{cc}
\frac{1}{2}(1-{\beta}^{2})&\beta\leq\frac{\sqrt{2}}{2}\\
(1+\frac{1}{2}{\beta}^{2}-\frac{1}{2}\sqrt{{\beta}^{4}+4{\beta}^{2}})(1-{\beta}^{2})&\beta\geq\frac{\sqrt{2}}{2}
\end{array}
\end{equation}

\textbf{Remark:} It should be pointed out that there are some
differences between the actual  success probability and  optimum
success probabilities for  the worst case of $\eta_{1}$. The former
depends on both of $\beta$ and  $\eta_{1}$, but the latter only
depends on $\beta$.

\subsection{Optimal unambiguous discrimination problems for Cases B4}

This subsection  discusses the optimal unambiguous discrimination
problem where complete classical knowledge of discriminated  states
but none   \emph{a priori} probabilities
 of preparing the discriminated states are available.

Here we have the same classical knowledge of discriminated states in
this case as in Section II. C,  thus we can follow the analysis in
Section II. C and choose ${\Pi}_{1}$ and ${\Pi}_{2}$ as Eqs.
(\ref{c1a}) and (\ref{c1b}).

In order to assure that ${\Pi}_{1}$, ${\Pi}_{2}$ and
${\Pi}_{0}=I-{\Pi}_{1}-{\Pi}_{2}$ be semi-positive, the
 constraints  on $\lambda_{1}$ and $\lambda_{2}$ given by Eqs. (\ref{c3}) and (\ref{c4}) should be satisfied
 where $|\langle\psi_{1}|\psi_{2}\rangle|=\beta$.

And what we shall do here is the same, i.e., to choose $\lambda_{1}$
and $\lambda_{2}$ based on \emph{a priori} information such that the
average success probability given by Eq. (\ref{c2}) is maximized
with the
 constraints in Eqs. (\ref{c3}) and (\ref{c4}).

When we have no knowledge of preparing probability $\eta_{1}$, our
task is to  choose $\lambda^{2,w\eta_{1}}_{1}(\beta)$ and
$\lambda^{2,w\eta_{1}}_{2}(\beta)$ to optimize the following
performance
\begin{equation}
\label{wc2}
J=\max\min_{\{\eta_{1}\}}[\lambda_{1}\eta_{1}+\lambda_{2}(1-\eta_{1})](1-\beta^{2})
\end{equation}
with the constraints described by Eqs. (\ref{c3}) and (\ref{c4}).

In this case, we have
\begin{equation}
\label{W28}
\lambda^{2,w\eta_{1}}_{1}(\beta)=\lambda^{2,w\eta_{1}}_{2}(\beta)=\frac{1}{1+\beta}
\end{equation}
and
\begin{equation}
\label{W28-1} P_{2}^{w\eta_{1}}(\beta)=1-\beta
\end{equation}
where the subscript $2$ of $P_{2}^{w\eta_{1}}$ means that we  have
the classical knowledge of both discriminated states, and the
superscript $w\eta_{1}$ implies that the  optimum success
probability is defined in terms of the worst case for $\eta_{1}$.

\textbf{Remark:} In this case, both the optimal decision given by
Eq. (\ref{W28}) and the optimum success probability  Eq.
(\ref{W28-1}) are just the functions of $\beta$.

\section{Discussion}

In this section, we further  investigate the role that \emph{a
priori} knowledge plays in quantum decision theory by analyzing the
effect of \emph{a priori} knowledge on  unambiguously discriminating
quantum states.

In general, the problems of unambiguously  discriminating quantum
states can be described in the language of  decision theory. As
mentioned in Ref. \cite{decision}, the key element of decision
theory is the loss function $L(\theta,\lambda)$  defined on the
other two elements, the parameter space $\Theta$ and the action
space $\Lambda$ ($\theta\in\Theta,\lambda\in\Lambda$).

Subsequently, we will not only examine the effect of \emph{a priori}
knowledge on the loss function (optimization), but also investigate
the influence of \emph{a priori} knowledge on both the parameter
space and the action space.

To begin with, \emph{a priori} knowledge for the aforementioned
eight cases is hereby addressed in Table \ref{tab.1}.

\begin{table}
\caption{\emph{a priori} information for eight cases.} \label{tab.1}
\begin{center}
\begin{tabular}{lrrrr}
Case   & $|\psi_{1}\rangle$ & $|\psi_{2}\rangle$ & $\beta$ &$\eta_{1}$\\
A1     & unknown    & unknown    & unknown & known    \\
B1     & unknown    & unknown    & unknown & unknown  \\
A2     & known      & unknown    & unknown & known    \\
B2     & known      & unknown    & unknown & unknown  \\
A3     & known      & unknown    & known   & known    \\
B3     & known      & unknown    & known   & unknown  \\
A4     & known      & known      & known   & known    \\
B4     & known      & known      & known   & unknown
\end{tabular}
\end{center}
\end{table}

\subsection{The role of \emph{a priori} knowledge in decision theory}

First, we will analyze the optimum unambiguous discrimination
problems in which none \emph{a priori} classical knowledge of
discriminated states is available.

\begin{table}
\caption{Comparing Case A1 with B1.} \label{tab.2}
\begin{center}
\begin{tabular}{lcc}
Case       &A1                  &B1                 \\
Parameter space $\Theta$     &$\beta,\eta_{1}$    &$\beta,\eta_{1}$   \\
Action  space $\Lambda$    & $\lambda_{1},\lambda_{2}$ & $\lambda_{1},\lambda_{2}$ \\
$P(\Theta,\Lambda)$  & Eq.(\ref{A8})    & Eq.(\ref{A8})  \\
Constraints on  $\Lambda$   & Eq.(\ref{A7})        & Eq.(\ref{A7})     \\
Loss function $L(\theta,\lambda)$  & Eq.(\ref{A8-J})    &Eq.(\ref{wc2}) \\
Action     & Eqs. (\ref{A-9a}-\ref{A-9b})      & Eq. (\ref{A-10a})  \\
Optimal performance   & Eq. (\ref{A-9})     & Eq. (\ref{A-11})
\end{tabular}
\end{center}
\end{table}

From Table \ref{tab.2}, we  see the same parameter space and action
space in Case A1 as in Case B1. Further more, there are also the
common expression of average success probability $P(\theta,\lambda)$
and the constraints on action space in both cases. However, the loss
function in Case B1 is quite different from that in Case A1, because
\emph{a priori} probability $\eta_{1}$ is available in Case A1 but
not in Case B1. Due to different \emph{a priori} knowledge in the
two cases, one has to make different decisions and take different
actions: one can take actions based on the knowledge of \emph{a
priori} probability $\eta_{1}$ in Case A1 but make decision without
the knowledge of $\eta_{1}$ in Case B1.

Next, we will analyze the optimum unambiguous discrimination
problems in which  the partial classical knowledge of the
discriminated states is provided.

\begin{table}
\caption{comparing Case A2 with B2.} \label{tab.3}
\begin{center}
\begin{tabular}{lcc}
Case       &A2                  &B2                 \\
Parameter space $\Theta$     &$\beta,\eta_{1}$    &$\beta,\eta_{1}$   \\
Action space $\Lambda$    & $\lambda_{1},\lambda_{2},\lambda_{3}$ & $\lambda_{1},\lambda_{2},\lambda_{3}$ \\
$P(\Theta,\Lambda)$  & Eq.(\ref{10})    & Eq.(\ref{10})  \\
Constraints on $\Lambda$    & Eqs.(\ref{i1}-\ref{i2})        & Eq.(\ref{i1}-\ref{i2})  \\
Loss function $L(\theta,\lambda)$  & Eq.(\ref{H10-1})    &Eq.(\ref{BH10-1})  \\
Action     & Eqs. (\ref{24-a}-\ref{24-c})      & Eqs.(\ref{BH10a}-\ref{BH10c})  \\
Optimal performance   & Eq. (\ref{24-1})      & Eq. (\ref{W24-1})
\end{tabular}
\end{center}
\end{table}

\begin{table}
\caption{comparing Case A3 with B3.} \label{tab.4}
\begin{center}
\begin{tabular}{lcc}
Case       &A3                  &B3                 \\
Parameter space $\Theta$     &$\beta,\eta_{1}$    &$\beta,\eta_{1}$   \\
Action  space $\Lambda$    & $\lambda_{1},\lambda_{2},\lambda_{3}$ & $\lambda_{1},\lambda_{2},\lambda_{3}$ \\
$P(\Theta,\Lambda)$  & Eq.(\ref{10})    & Eq.(\ref{10})  \\
Constraints on  $\Lambda$   & Eqs.(\ref{i1}-\ref{i2})        & Eq.(\ref{i1}-\ref{i2})  \\
Loss function $L(\theta,\lambda)$  & Eq.(\ref{A10})    &Eq.(\ref{10-1})  \\
Action     & Eqs. (\ref{A28a}-\ref{A28c})      & Eqs.(\ref{wA28a}-\ref{wA28c})  \\
Optimal performance   & Eq.(\ref{A28-1})      & Eq. (\ref{wA28-2})
\end{tabular}
\end{center}
\end{table}

From both Tables \ref{tab.3} and \ref{tab.4}, we have found that the
average success probability functions share the same expressions in
Cases A2, B2, A3 and B3. In addition, both the parameter spaces and
the action spaces are the same in these four cases. The same
constraints on the action spaces must be satisfied in the four
cases. However, the loss functions in the four cases are quite
different. Due to different \emph{a priori} knowledge in the four
cases, we have to make different decisions: one can take action with
 both \emph{a priori} probability $\eta_{1}$ and \emph{a priori}
knowledge of $\beta$ in Case A3, and only with the knowledge of
$\eta_{1}$ in Case A2; one can make decision only based on \emph{a
priori} knowledge of $\beta$ in Case B3,  and  without any knowledge
of $\eta_{1}$ and  $\beta$ in Case B2.

Finally, we will analyze the cases of the optimum unambiguous
discrimination problems provided with  the complete classical
knowledge of the discriminated states.

\begin{table}
\caption{comparing Case A4 with B4.} \label{tab.5}
\begin{center}
\begin{tabular}{lcc}
Case       &A4                 &B4                 \\
Parameter space $\Theta$     &$\beta,\eta_{1}$    &$\beta,\eta_{1}$   \\
Action  space $\Lambda$    & $\lambda_{1},\lambda_{2}$ & $\lambda_{1},\lambda_{2}$ \\
$P(\Theta,\Lambda)$  & Eq.(\ref{c2})    & Eq.(\ref{c2})  \\
Constraints on $\Lambda$    & Eqs.(\ref{c3}-\ref{c4})        & Eq.(\ref{c3}-\ref{c4})  \\
Loss function $L(\theta,\lambda)$  & Eq.(\ref{c2-1})    &Eq.(\ref{wc2})  \\
Action     & Eqs. (\ref{28a}-\ref{28b})      & Eqs.(\ref{W28})  \\
Optimal performance   & Eq.(\ref{28-1})      & Eq. (\ref{W28-1})
\end{tabular}
\end{center}
\end{table}

From  Table \ref{tab.5}, one can find that the average success
probability functions have the same expression in both Cases A4 and
B4. In addition, both the parameter spaces and the action spaces are
the same in both cases. The  constrained conditions on the action
spaces must be satisfied in the four cases. However, the loss
functions in the both cases are different. Due to different \emph{a
priori} information in the four cases, we have to make different
decisions: one can take action with
 both \emph{a priori} probability $\eta_{1}$ and \emph{a priori}
knowledge of $\beta$ in Case A4, and only with \emph{a priori}
knowledge of $\beta$ in Case B4.

\subsection{The effect of \emph{a priori} information on optimum performances}

\begin{figure}[ht]
\centering \subfigure[$P_{0}^{opt}(\beta,\eta_{1})$] {
\label{Fig.1:a} %% label for first subfigure
 \scalebox{0.22}{\includegraphics{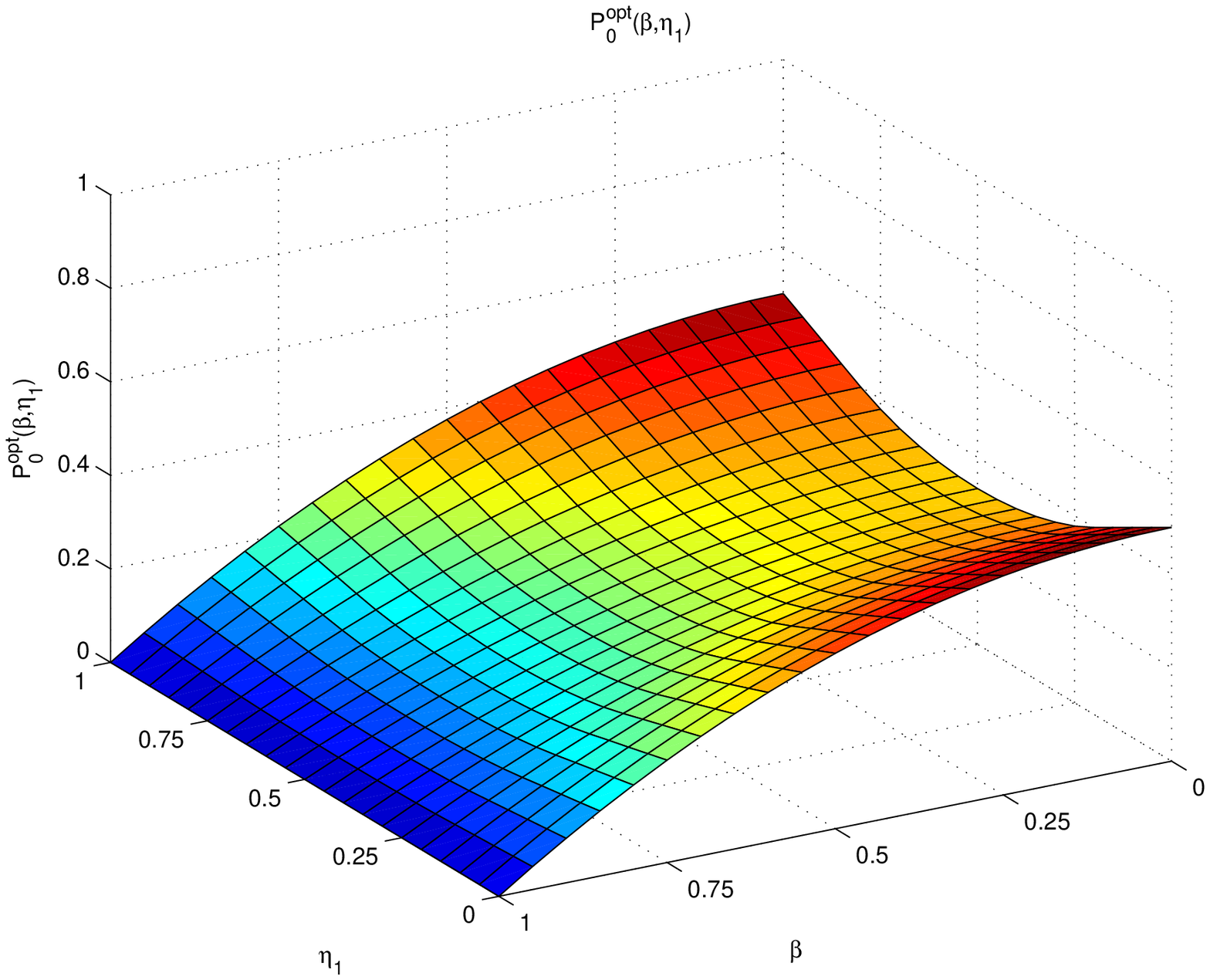}}}%
\subfigure[$P_{1}^{w\beta}(\beta,\eta_{1})$] {
\label{Fig.1:b} %% label for second subfigure
 \scalebox{0.22}{\includegraphics{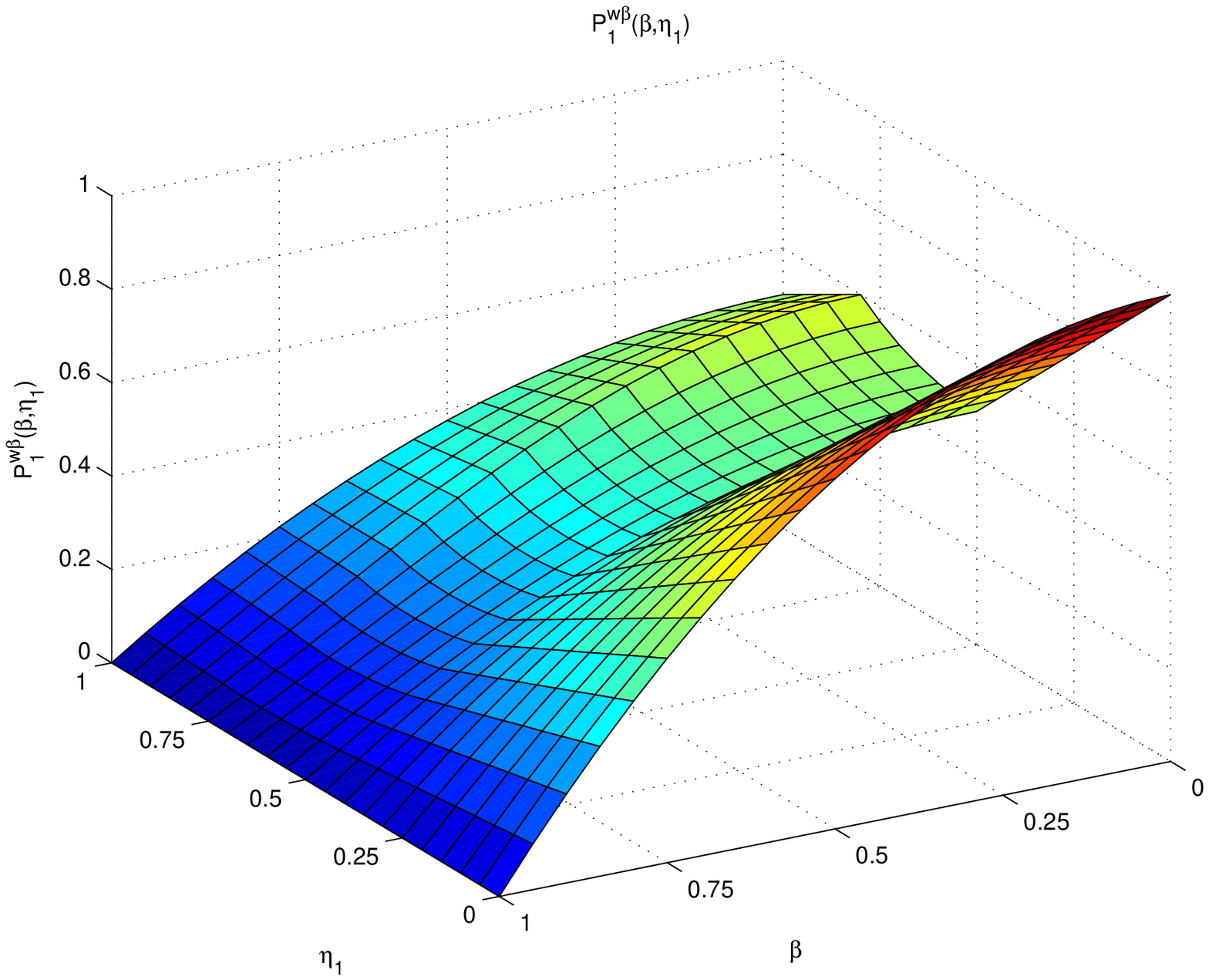}}
}%
\\
\centering \subfigure[$P_{1+}^{opt}(\beta,\eta_{1})$] {
 \label{Fig.1:c} %% label for third subfigure
\scalebox{0.22}{\includegraphics{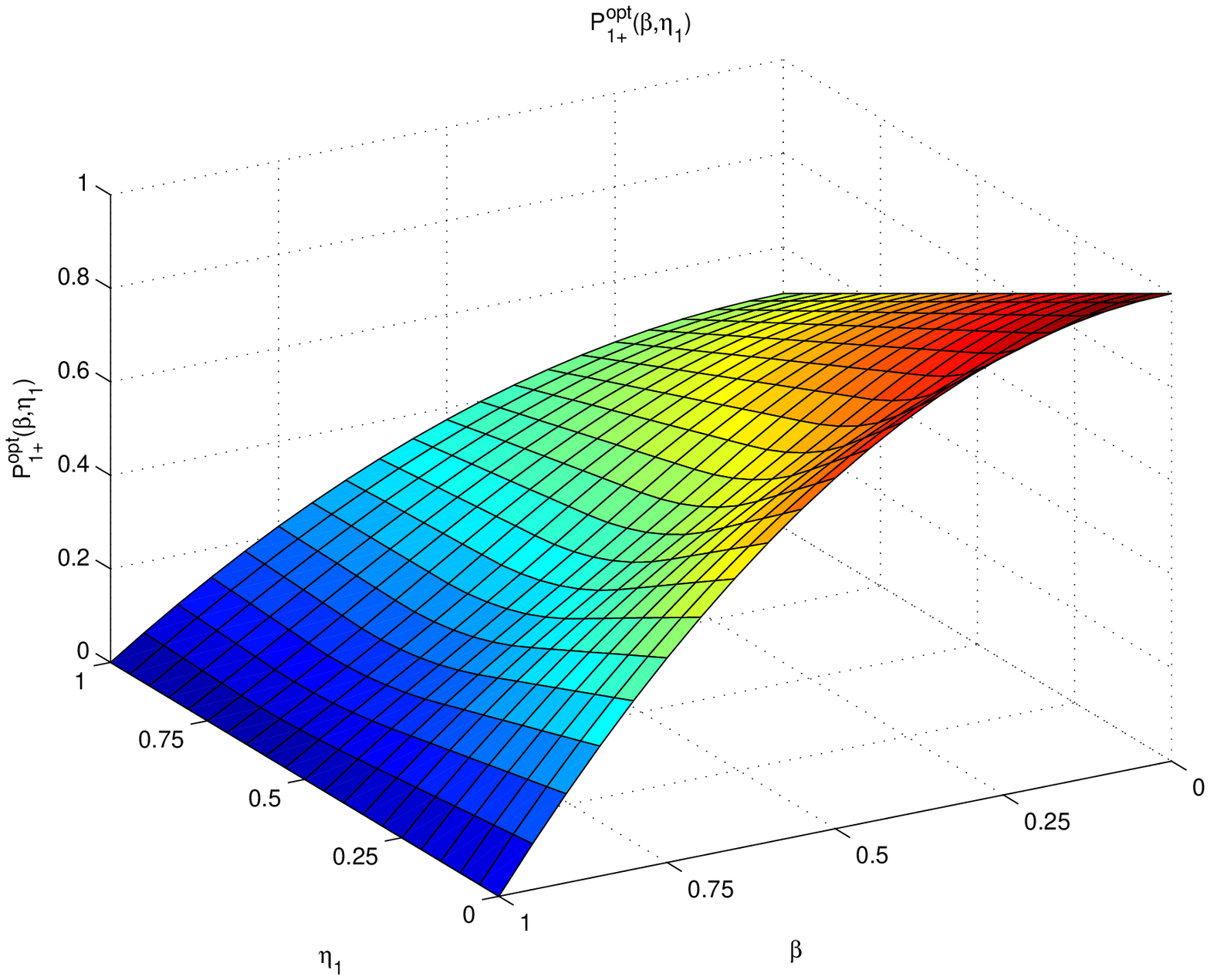}}
}%
\subfigure[$P_{2}^{opt}(\beta,\eta_{1})$] {
 \label{Fig.1:d} %% label for fourth subfigure
\scalebox{0.22}{\includegraphics{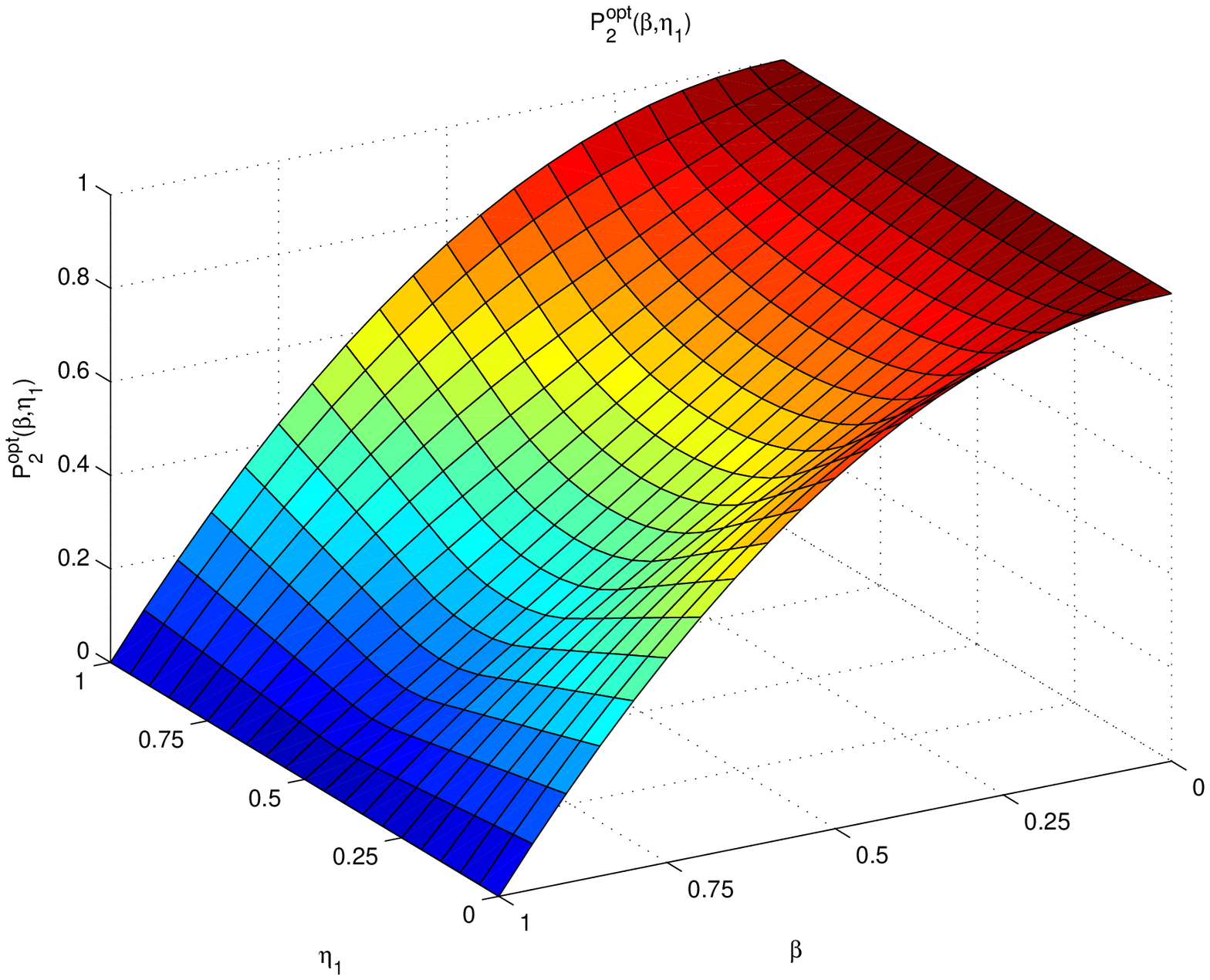}}
}%
\caption{\label{fig1} Optimum success probabilities:
$P_{0}^{opt}(\beta,\eta_{1})$, $P_{1}^{w\beta}(\beta,\eta_{1})$,
$P_{1+}^{opt}(\beta,\eta_{1})$ and
$P_{2}^{opt}(\beta,\eta_{1})$} %% label for entire figure
\end{figure}

\begin{figure}
\centering
\scalebox{0.29}{\includegraphics{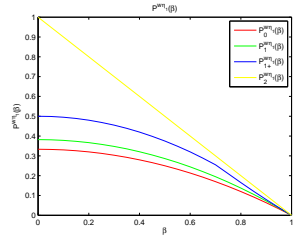}} %100 percent
\vspace*{13pt}% Here is how to import EPS art
\caption{\label{fig2} $P_{0}^{w\eta_{1}}(\beta)$,
$P_{1}^{w\beta\eta_{1}}(\beta)$, $P_{1+}^{w\eta_{1}}(\beta)$ and
$P_{2}^{w\eta_{1}}(\beta)$ and  their images.}
\end{figure}

To carry out comparative analysis, we  plot
$P_{0}^{opt}(\beta,\eta_{1})$, $P_{1}^{w\beta}(\beta,\eta_{1})$,
$P_{1+}^{opt}(\beta,\eta_{1})$ and $P_{2}^{opt}(\beta,\eta_{1})$ in
Fig.\ref{fig1} and plot $P_{0}^{w\eta_{1}}(\beta)$,
$P_{1}^{w\beta\eta_{1}}(\beta)$, $P_{1+}^{w\eta_{1}}(\beta)$ and
$P_{2}^{w\eta_{1}}(\beta)$ in Fig. \ref{fig2}.

\begin{figure}[ht]
\centering \subfigure[$P_{0}^{w\eta_{1}}(\beta,\eta_{1})$] {
\label{Fig.3:a} %% label for first subfigure
 \scalebox{0.22}{\includegraphics{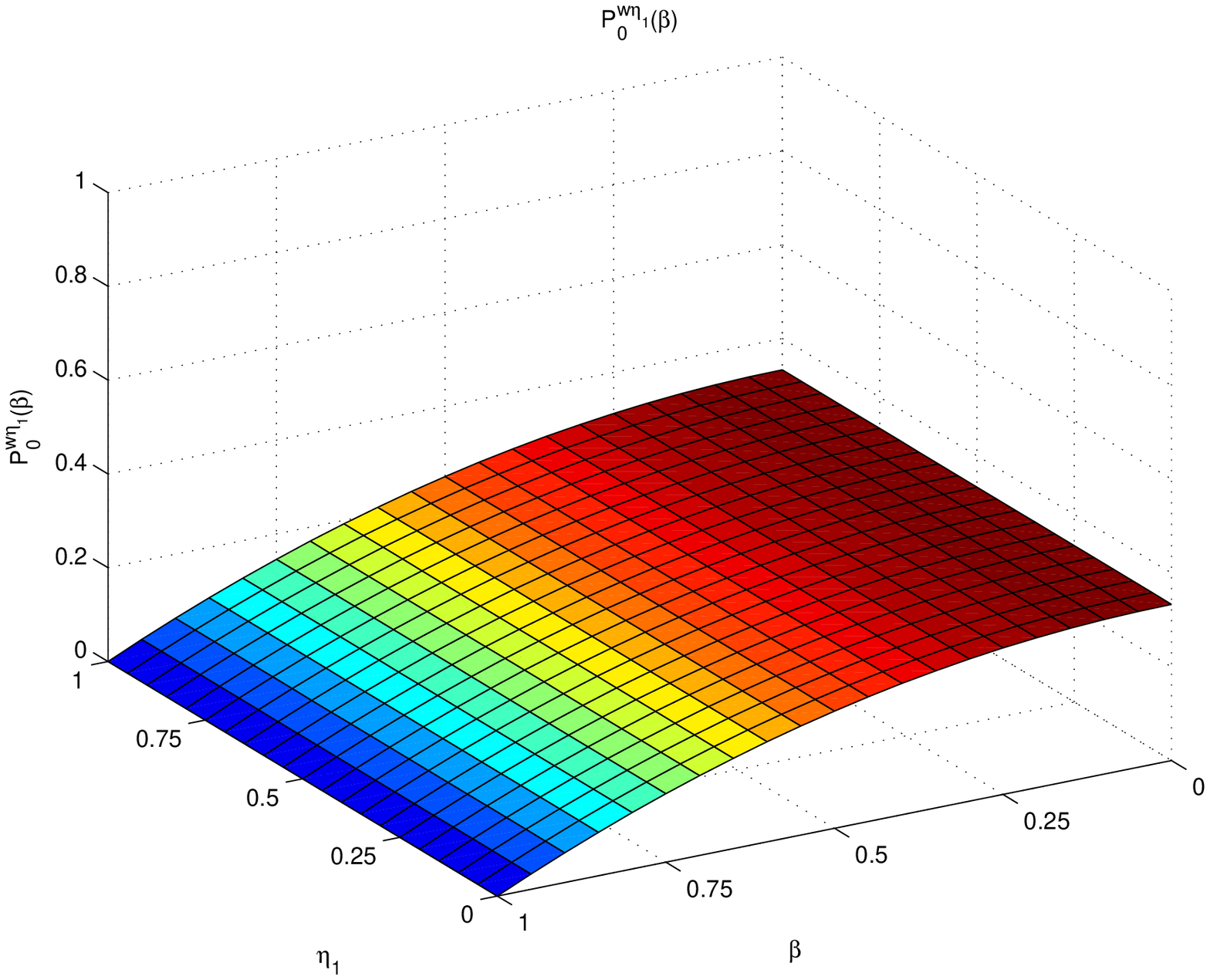}}
}%
\subfigure[$P_{1}^{w\beta\eta_{1}}(\beta,\eta_{1})$] {
\label{Fig.3:b} %% label for second subfigure
 \scalebox{0.22}{\includegraphics{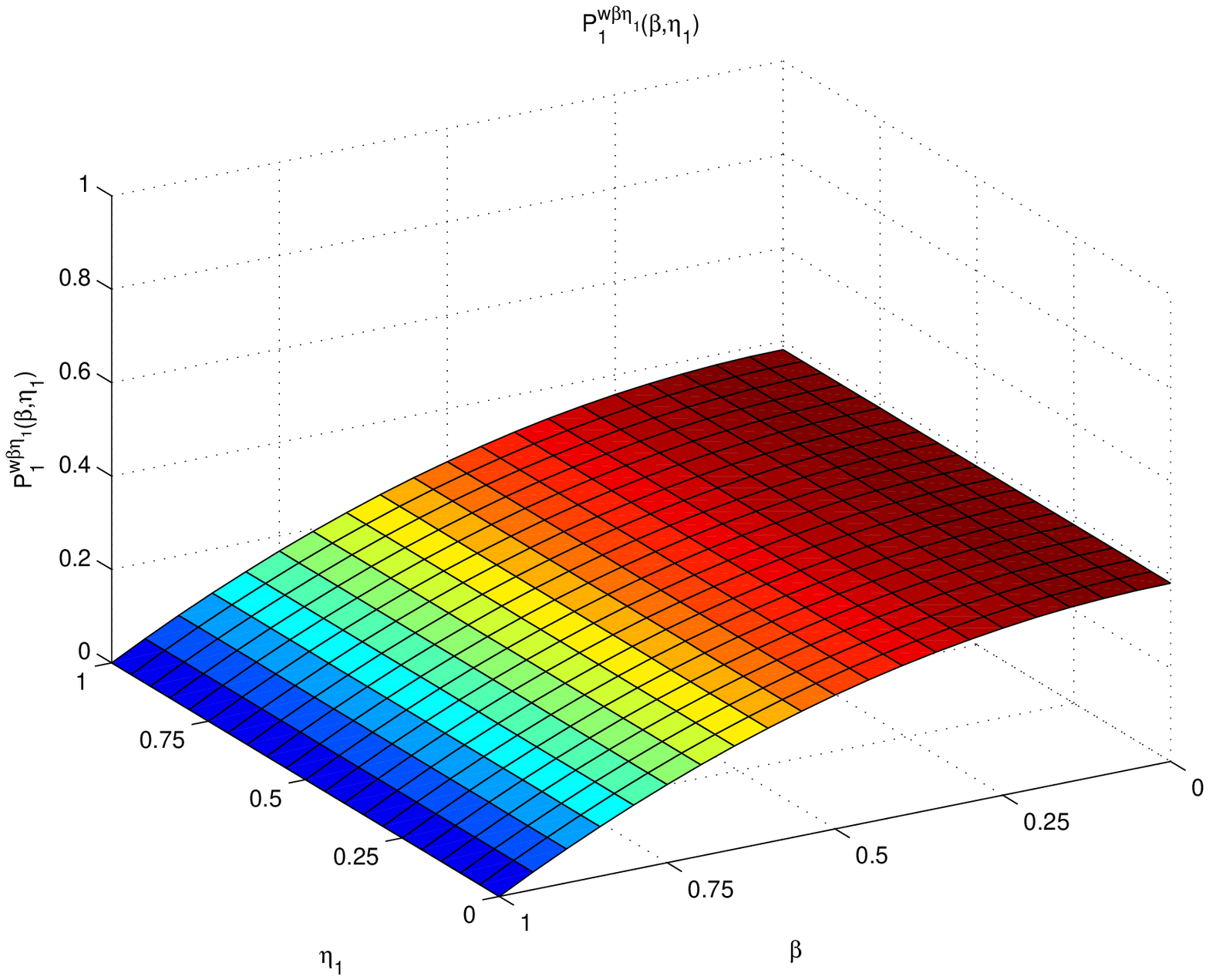}}
}%
\\
\centering \subfigure[$P_{1+}^{w\eta_{1}}(\beta,\eta_{1})$] {
 \label{Fig.3:c} %% label for third subfigure
\scalebox{0.22}{\includegraphics{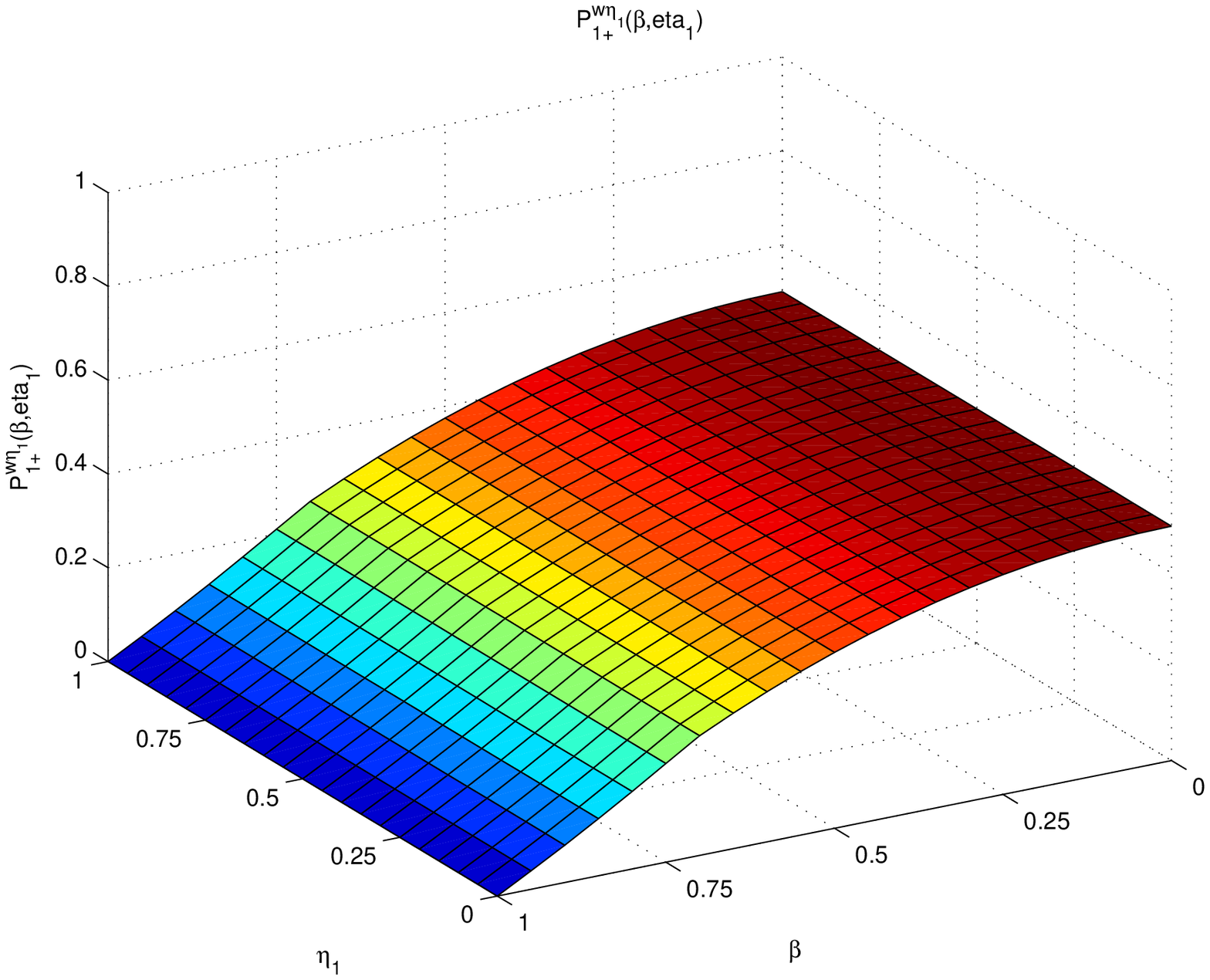}}
}%
\subfigure[$P_{2}^{w\eta_{1}}(\beta,\eta_{1})$] {
 \label{Fig.3:d} %% label for fourth subfigure
\scalebox{0.22}{\includegraphics{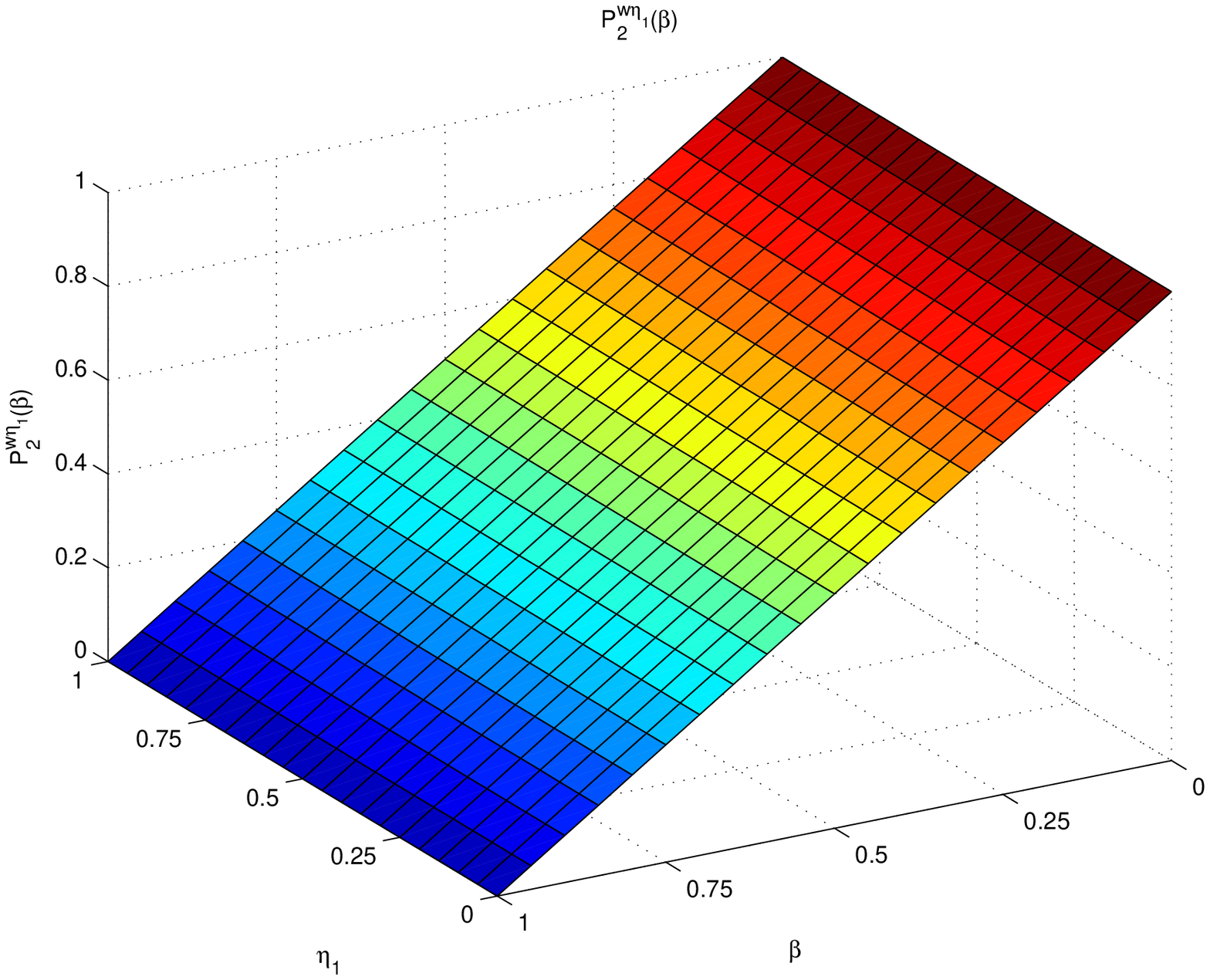}}
}%
\caption{\label{fig3} Optimum success probabilities:
$P_{0}^{w\eta_{1}}(\beta,\eta_{1})$,
$P_{1}^{w\beta\eta_{1}}(\beta,\eta_{1})$,
$P_{1+}^{w\eta_{1}}(\beta,\eta_{1})$ and
$P_{2}^{w\eta_{1}}(\beta,\eta_{1})$} %% label for entire figure
\end{figure}

To make even clearer comparison, we further plot two-parameter
functions $P_{0}^{w\eta_{1}}(\beta,\eta_{1})$,
$P_{1}^{w\beta\eta_{1}}(\beta,\eta_{1})$,
$P_{1+}^{w\eta_{1}}(\beta,\eta_{1})$ and
$P_{2}^{w\eta_{1}}(\beta,\eta_{1})$ in Fig. \ref{fig3}.

\subsubsection{The effect of \emph{a priori}  probabilities on
optimum performances}

\begin{figure}[ht]
\centering
\subfigure[$P_{0}^{w\eta_{1}\rightarrow\eta_{1}}(\beta,\eta_{1})$] {
\label{Fig.4:a} %% label for first subfigure
 \scalebox{0.22}{\includegraphics{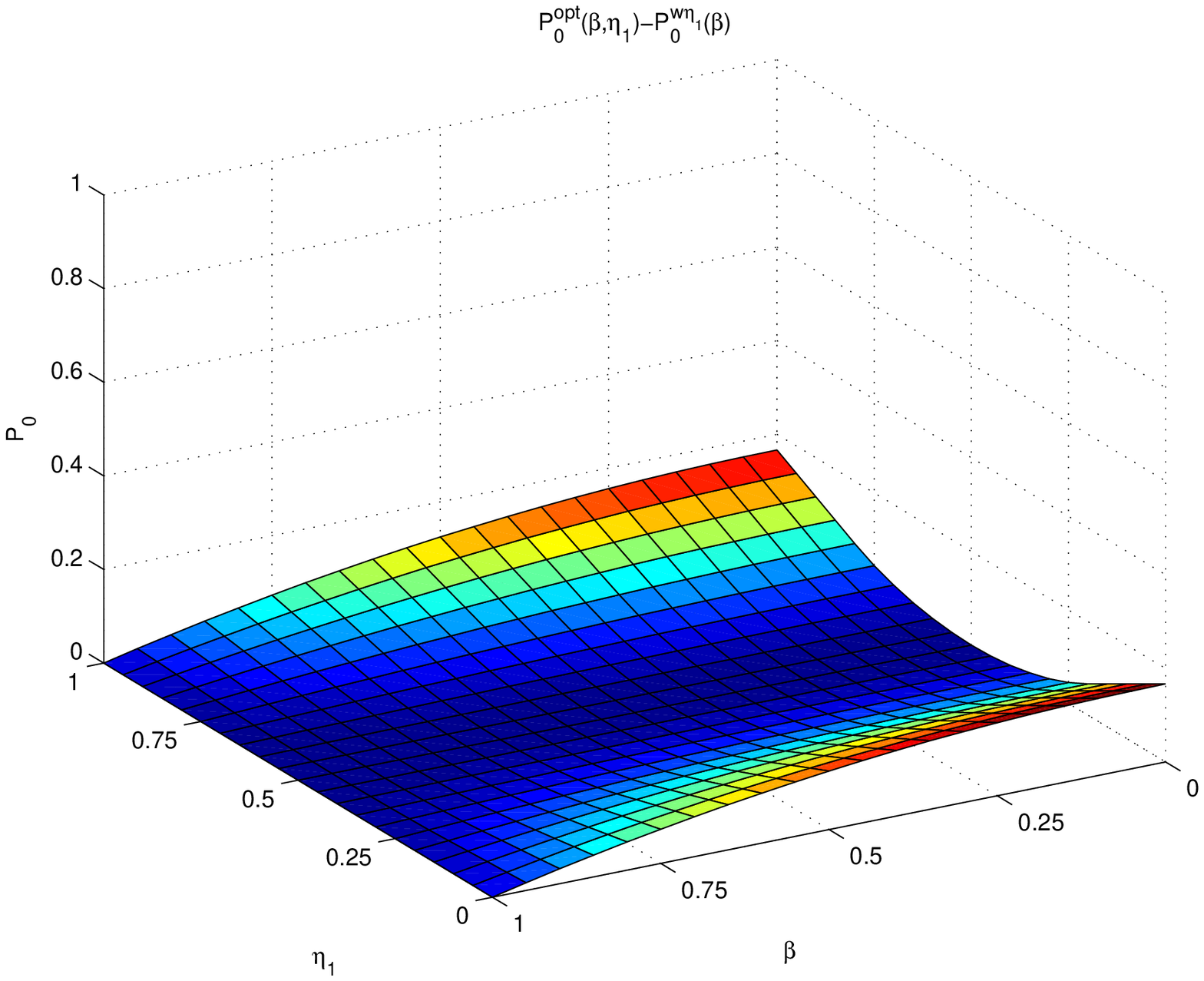}}
}%
\subfigure[$P_{1}^{w\eta_{1}\rightarrow\eta_{1}}(\beta,\eta_{1})$] {
\label{Fig.4:b} %% label for second subfigure
 \scalebox{0.22}{\includegraphics{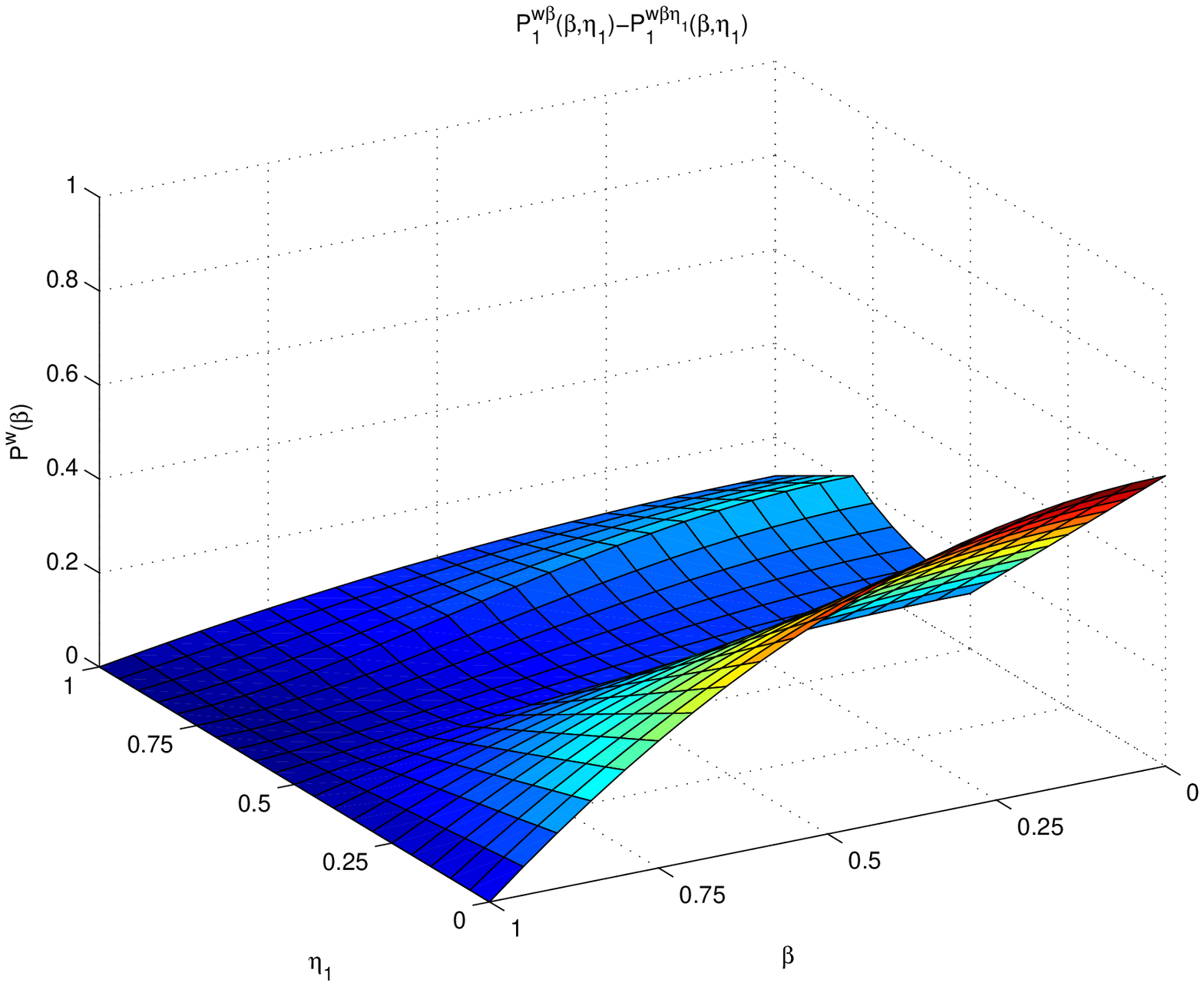}}
}%
\\
\centering
\subfigure[$P_{1+}^{w\eta_{1}\rightarrow\eta_{1}}(\beta,\eta_{1})$]
{
 \label{Fig.4:c} %% label for third subfigure
\scalebox{0.22}{\includegraphics{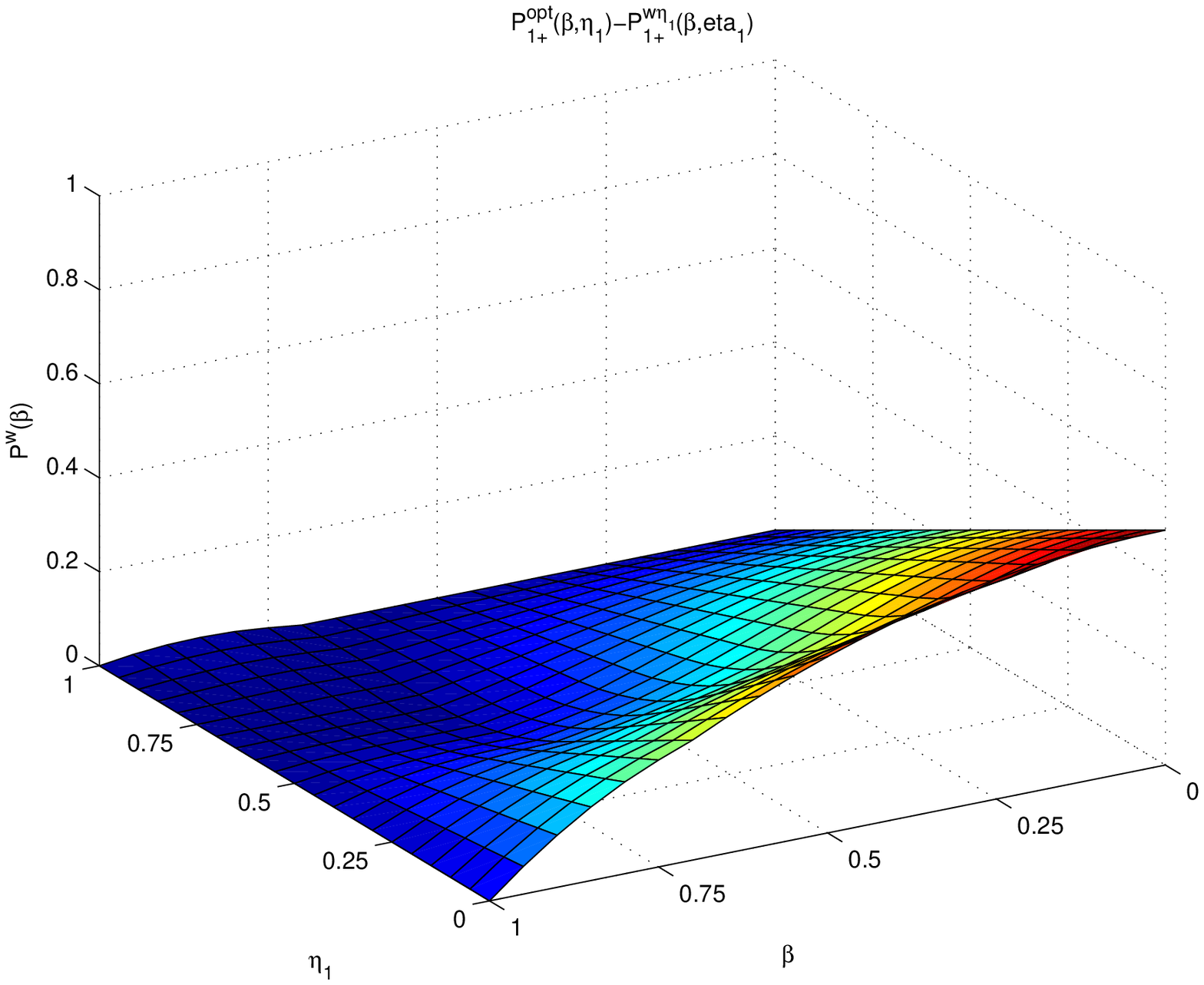}}
}%
\subfigure[$P_{2}^{w\eta_{1}\rightarrow\eta_{1}}(\beta,\eta_{1})$] {
 \label{Fig.4:d} %% label for fourth subfigure
\scalebox{0.22}{\includegraphics{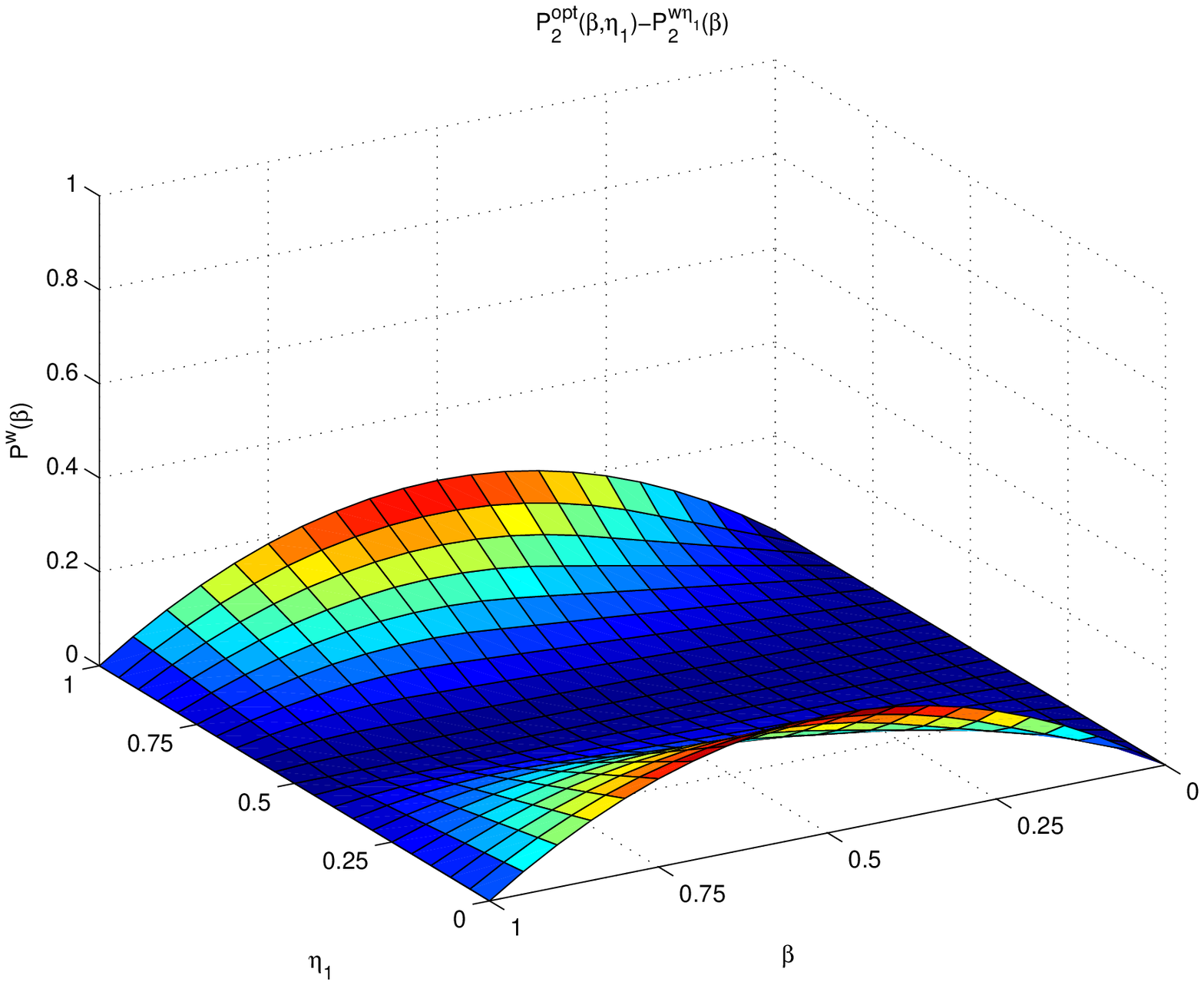}}
}%
\caption{\label{fig4} Improving optimum success probabilities:
$P_{0}^{w\eta_{1}\rightarrow\eta_{1}}(\beta,\eta_{1})$,
$P_{1}^{w\eta_{1}\rightarrow\eta_{1}}(\beta,\eta_{1})$,
$P_{1+}^{w\eta_{1}\rightarrow\eta_{1}}(\beta,\eta_{1})$ and
$P_{2}^{w\eta_{1}\rightarrow\eta_{1}}(\beta,\eta_{1})$} %% label for entire figure
\end{figure}

By comparing the optimal performance $P_{0}^{opt}(\beta,\eta_{1})$
(see Fig. \ref{Fig.1:a}) in Case A1 with
$P_{0}^{w\eta_{1}}(\beta,\eta_{1})$ (Fig. \ref{Fig.3:a}) in Case B1,
it is demonstrated that the former is better than the latter as
shown in Fig. \ref{Fig.4:a}. This implies that \emph{a priori}
probability can be utilized to improve the optimum performance even
when the two discriminated states are classically unknown, yet we
have a copy of them.

By comparing the optimal performance
$P_{1}^{w\beta}(\beta,\eta_{1})$(Fig. \ref{Fig.1:b}) in Case A2 with
$P_{1}^{w\beta\eta_{1}}(\beta,\eta_{1})$ (see Fig. \ref{Fig.3:b}) in
Case B2, the former is found to be better than the latter as shown
in Fig. \ref{Fig.4:b}. It is also easy to find in Fig. \ref{Fig.4:c}
that the optimal performance $P_{1+}^{opt}(\beta,\eta_{1})$ (see
Fig. \ref{Fig.1:c})in Case A3 is better than
$P_{1+}^{w\eta_{1}}(\beta,\eta_{1})$  (see Fig. \ref{Fig.3:c}) in
Case B3. This implies that the \emph{a priori} probability can be
utilized to
 improve the optimum performance when one only has partial
\emph{a priori} classical knowledge of the discriminated states.

In the comparing Figure \ref{Fig.4:d} between the optimal
performance $P_{2}^{opt}(\beta,\eta_{1})$ (see Fig. \ref{Fig.1:d})
in Case A4 and $P_{2}^{w\eta_{1}}(\beta,\eta_{1})$ (see Fig.
\ref{Fig.3:d}) in Case B4, the same conclusion can be addressed,
i.e., the former is clearly better than the latter. This implies
that the knowledge of \emph{a priori} probability can be utilized to
improve the optimum performance when we  have \emph{a priori}
complete classical knowledge of the discriminated states.

\subsubsection{The effect of classical knowledge on optimum
performances} 
\begin{figure}[ht]
\centering \subfigure[$P_{0\rightarrow1}^{opt}(\beta,\eta_{1})$]{
\label{Fig.5:a} %% label for first subfigure
\scalebox{0.22}{\includegraphics{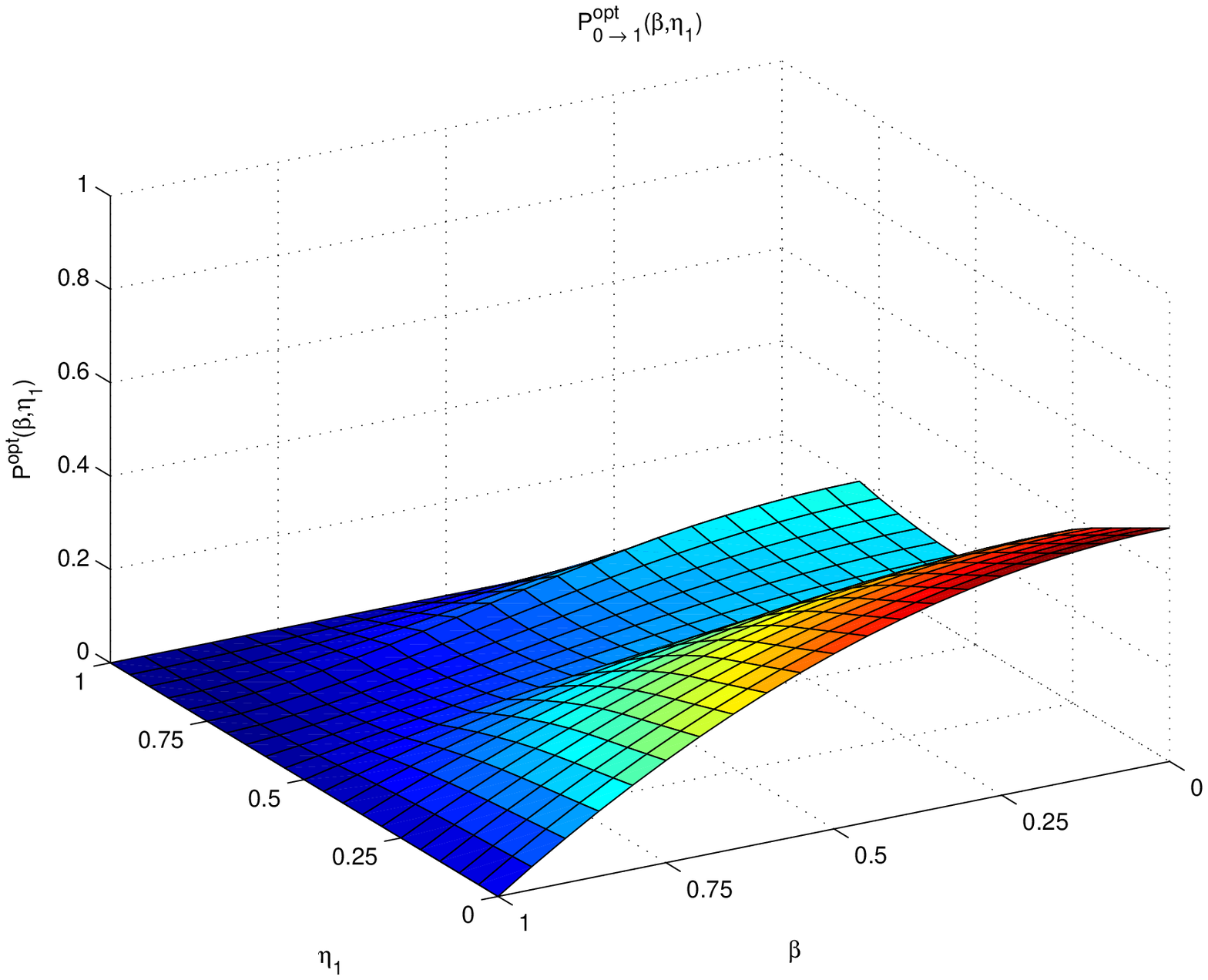}}
}%
\subfigure[$P_{1\rightarrow1+}^{opt}(\beta,\eta_{1})$] {
\label{Fig.5:b} %% label for second subfigure
 \scalebox{0.22}{\includegraphics{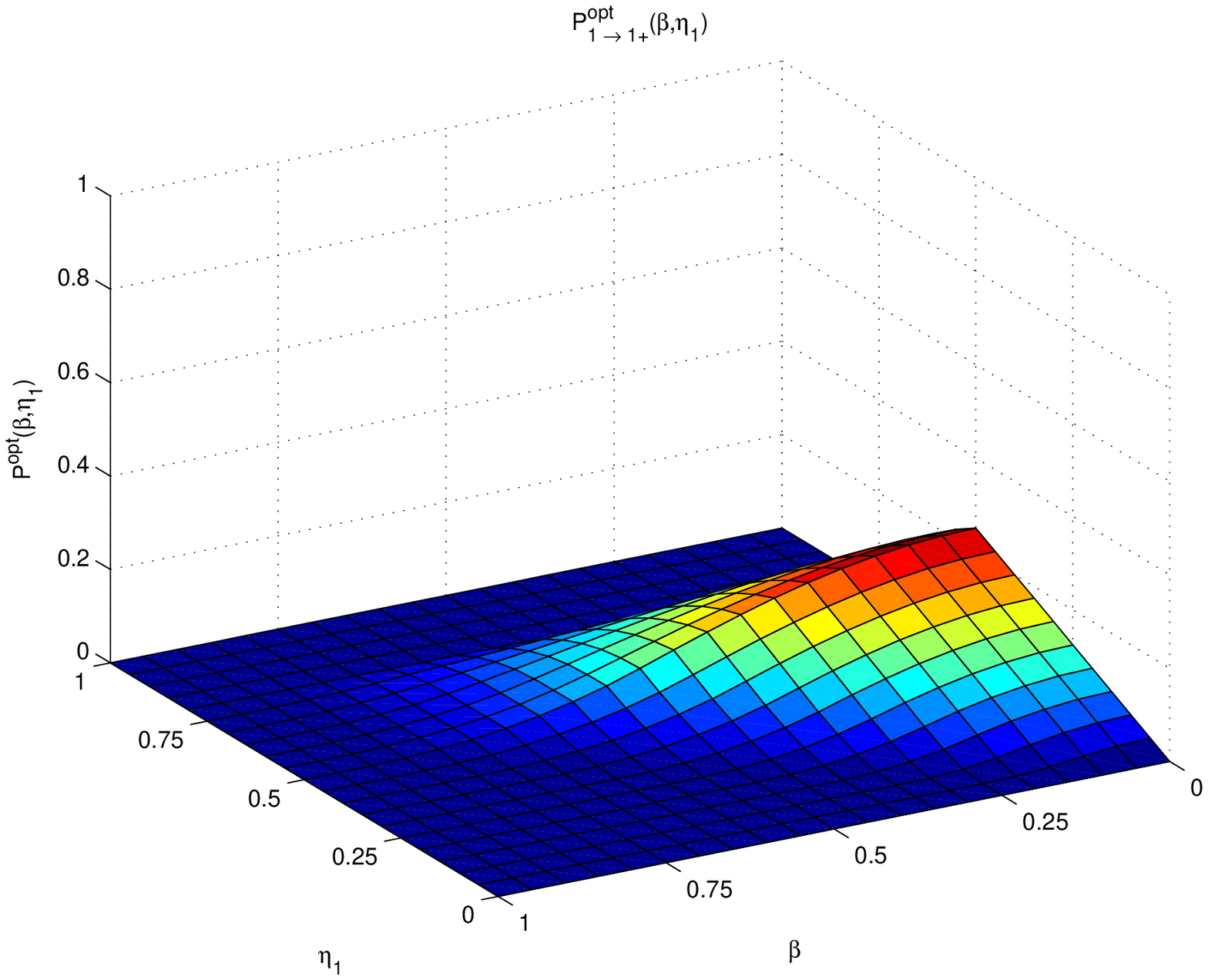}}
}%
\\
\subfigure[$P_{1+\rightarrow2}^{opt}(\beta,\eta_{1})$] {
\label{Fig.5:c} %% label for second subfigure
 \scalebox{0.22}{\includegraphics{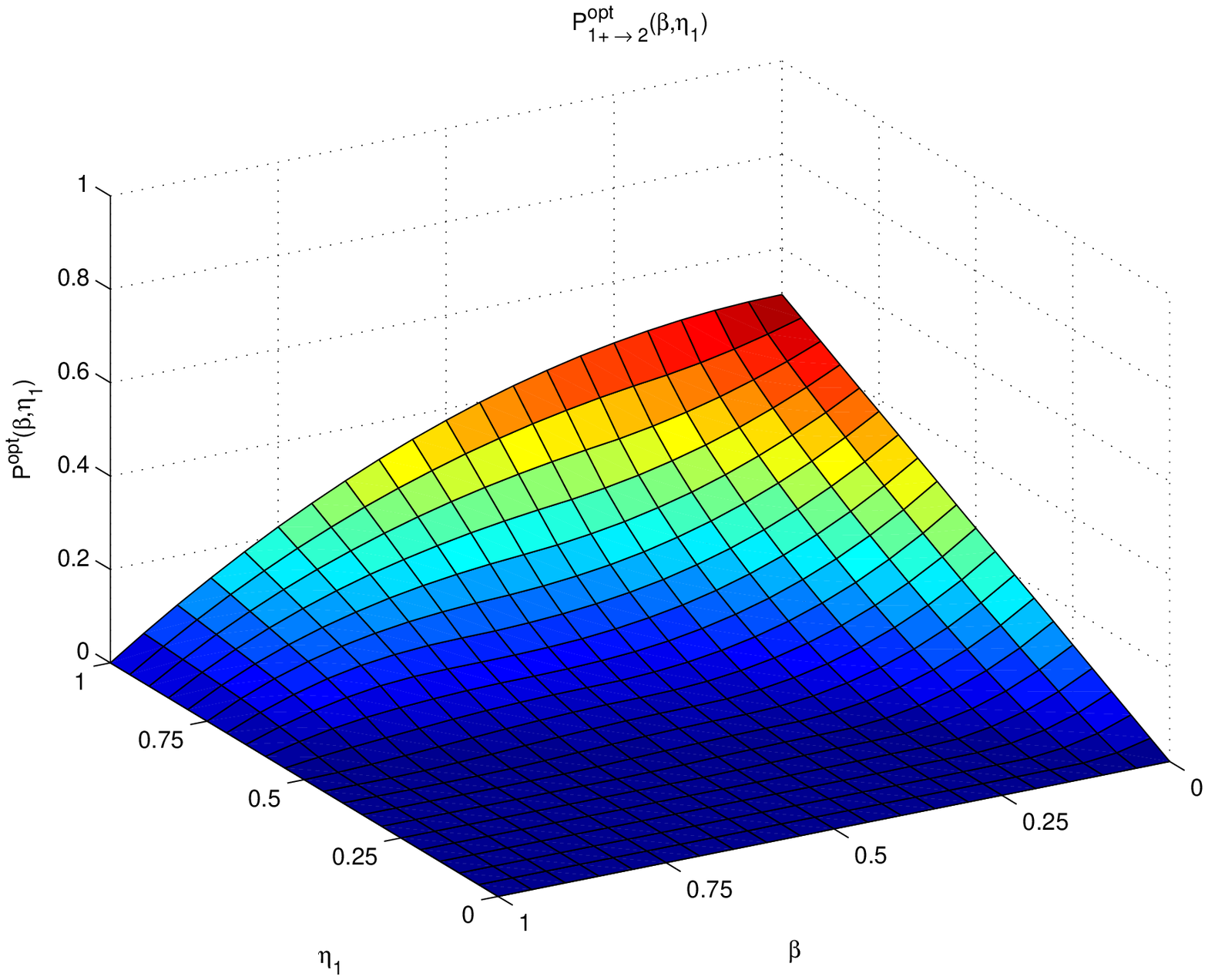}}
}%
\\
\caption{\label{fig5} Performance-improving functions
$P_{0\rightarrow1}^{opt}(\beta,\eta_{1})$,
$P_{1\rightarrow1+}^{opt}(\beta,\eta_{1})$ and
$P_{1+\rightarrow2}^{opt}(\beta,\eta_{1})$}
\end{figure}

As shown in Fig. \ref{Fig.5:a}, the optimal
performance $P_{0}^{opt}(\beta,\eta_{1})$ (see Fig. \ref{Fig.1:a})
in Case A1 is better than $P_{1}^{w\beta}(\beta,\eta_{1})$ (Fig.
\ref{Fig.1:b}) in Case A2, and naturally the optimal performance
$P_{1+}^{opt}(\beta,\eta_{1})$ (Fig. \ref{Fig.1:c}) in Case A3 is
better than $P_{1}^{w\beta}(\beta,\eta_{1})$ (Fig. \ref{Fig.1:b}) in
Case A2,  as shown in Fig. \ref{Fig.5:b}, the optimal performance
$P_{1+}^{opt}(\beta,\eta_{1})$ (Fig. \ref{Fig.1:c}) in Case A3 is
better than $P_{2}^{opt}(\beta,\eta_{1})$ (see Fig. \ref{Fig.1:d})
in Case A4,  as shown in Fig. \ref{Fig.5:c}. This implies that the
classical knowledge of discriminated states can be utilized to
improve the optimum performance when we  have \emph{a priori}
probability of preparing the discriminated states.

\begin{figure}[ht]
\centering
\subfigure[$P_{0\rightarrow1}^{w\eta_{1}}(\beta,\eta_{1})$]{
\label{Fig.6:a} %% label for first subfigure
\scalebox{0.22}{\includegraphics{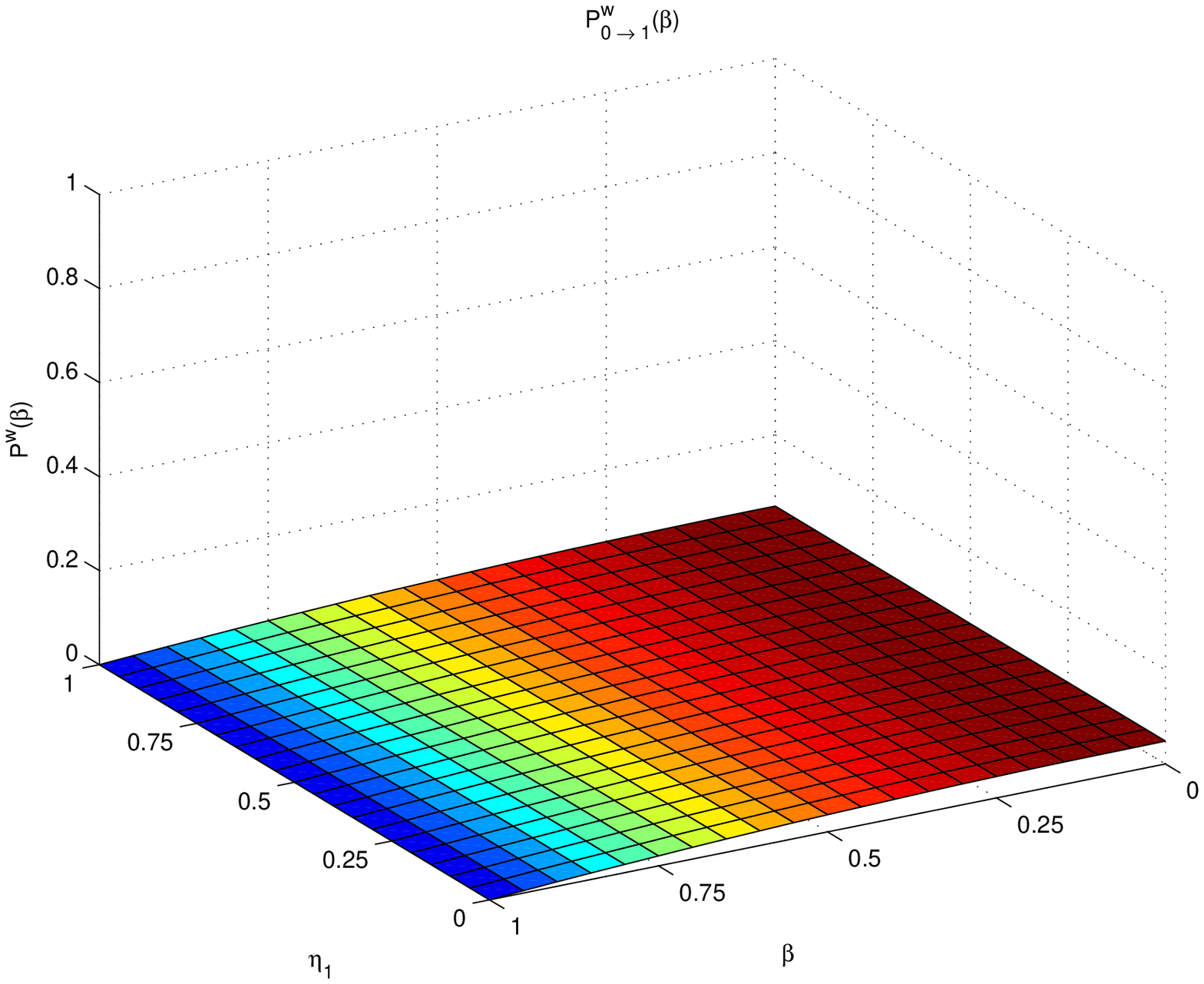}}
}%
\\
\subfigure[$P_{1\rightarrow1+}^{w\eta_{1}}(\beta,\eta_{1})$] {
\label{Fig.6:b} %% label for second subfigure
 \scalebox{0.22}{\includegraphics{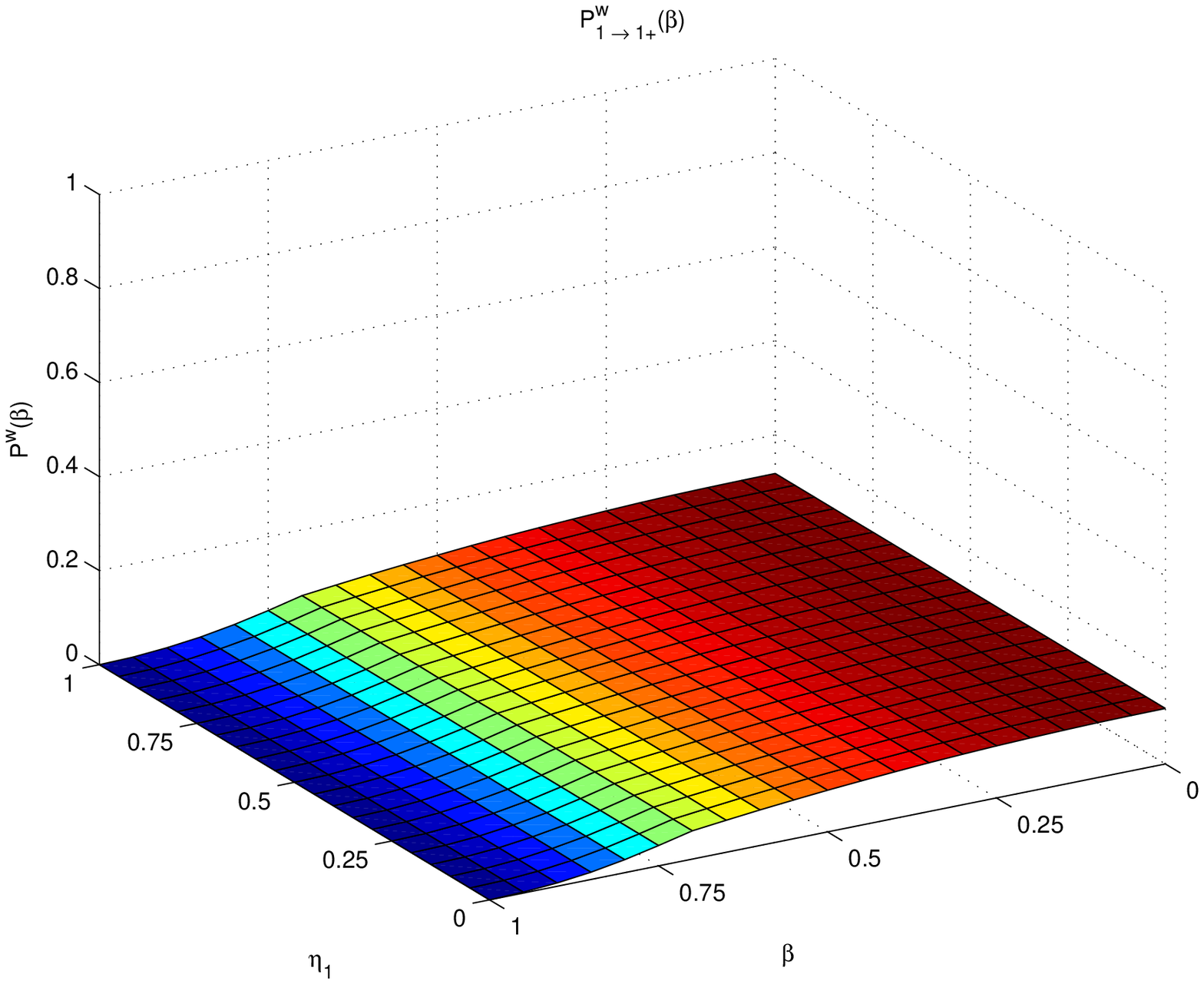}}
}%
\subfigure[$P_{1+\rightarrow2}^{w\eta_{1}}(\beta,\eta_{1})$] {
\label{Fig.6:c} %% label for second subfigure
 \scalebox{0.22}{\includegraphics{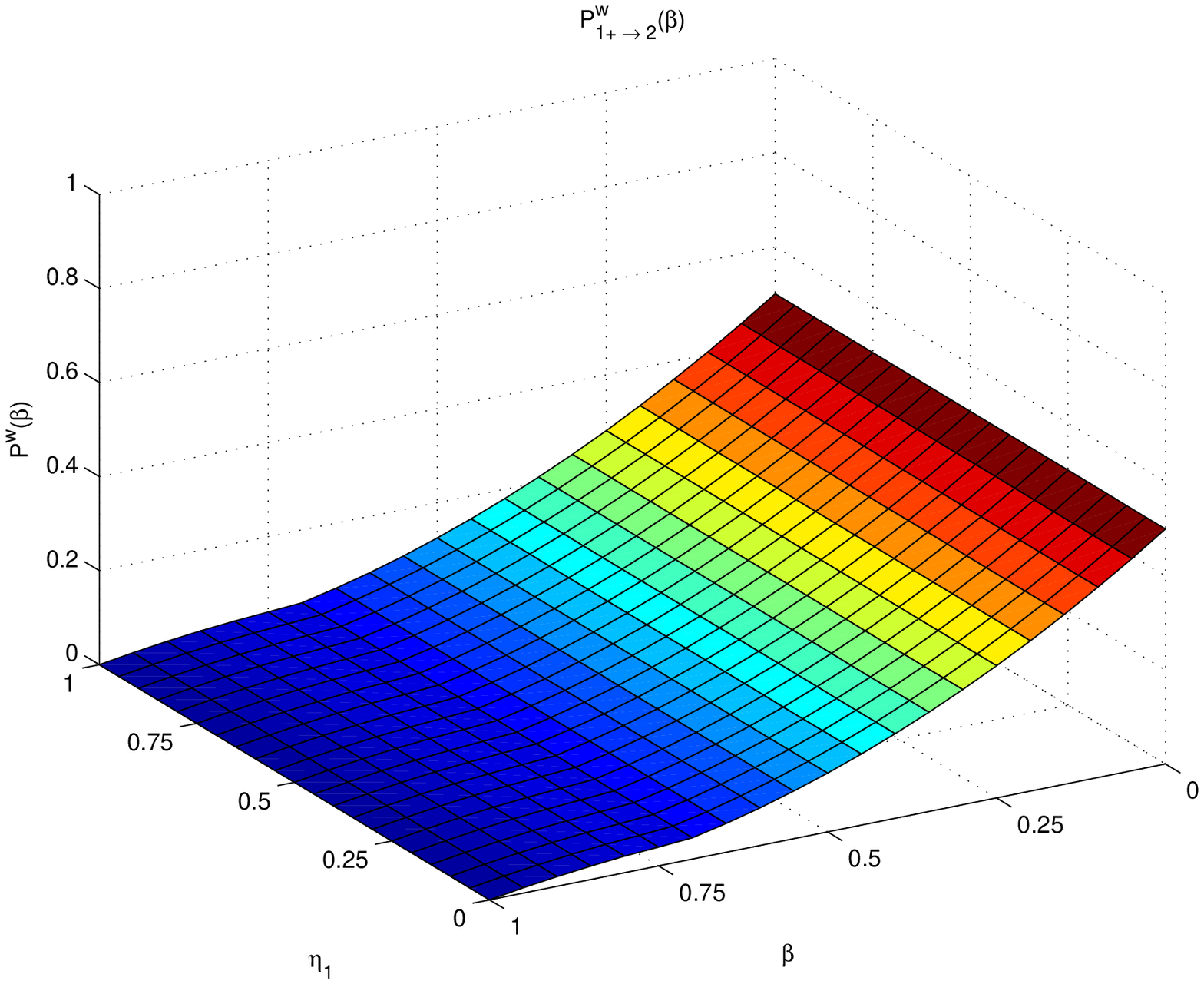}}
}%
\\
\caption{\label{fig6} Two-parameter performance-improving functions
$P_{0\rightarrow1}^{w\eta_{1}}(\beta,\eta_{1})$,
$P_{1\rightarrow1+}^{w\eta_{1}}(\beta,\eta_{1})$ and
$P_{1+\rightarrow2}^{w\eta_{1}}(\beta,\eta_{1})$}
\end{figure}

 It
is observed in Fig. \ref{fig6} that the optimum performance
increases with the amount of classical knowledge of discriminated
states provided even when
 \emph{a priori} probabilities of preparing the
discriminated states are unknown.

By comparing the optimal performance
$P_{0}^{w\eta_{1}}(\beta,\eta_{1})$ (see Fig. \ref{Fig.3:a})in Case
B1 with $P_{1}^{w\beta\eta_{1}}(\beta,\eta_{1})$ (Fig.
\ref{Fig.3:b}) in Case B2, the  latter is better than the  former as
shown in Fig. \ref{Fig.6:a}. It is not difficult to predict the same
situation in Fig. \ref{Fig.6:b} and  Fig.\ref{Fig.6:c} when
comparing the optimal performance
$P_{1+}^{w\eta_{1}}(\beta,\eta_{1})$ (Fig. \ref{Fig.3:c}) in Case B3
with $P_{1}^{w\beta\eta_{1}}(\beta,\eta_{1})$ (Fig. \ref{Fig.3:b})
in Case B2,  and when comparing the optimal performance
$P_{1+}^{w\eta_{1}}(\beta,\eta_{1})$ (see Fig. \ref{Fig.3:c}) in
Case B3 with $P_{2}^{w\eta_{1}}(\beta,\eta_{1})$ (see Fig.
\ref{Fig.3:d}) in Case B4.

\section{Optimal unambiguous discrimination problems for two qutrit states}

To further clarify the effect of \emph{a priori} information of the
discriminated states on optimum unambiguous discriminators
 and optimum success probabilities, we will  study   optimum   unambiguous discrimination problems
 for  two qutrit states in this section.
 At first, we will present the results on
 optimal
 unambiguous discrimination problems with  the knowledge of  \emph{a priori} preparing
 probabilities in the first three subsections.  According to what kind of classical knowledge  can be
 utilized,  the 4 {cases}
 are discussed as follows

Case A1, without classical knowledge of either state but
 with a single copy of  unknown states;

Case A2,  with only classical knowledge of one of the two states and
a single copy of the other unknown state;

Case A3, with only classical knowledge of one of the two states and
the absolute value of the inner product of both states, and also
with a single copy of the other unknown state;

Case A4,  with classical knowledge of both states.

The A1 and A4 cases will be investigated in subsection A and C,
respectively, and the A2 and A3 cases will be studied in subsection
B.

  Furthermore, optimal unambiguous discrimination problems
without \emph{a priori} preparing
 probability will be investigated in the subsection D, E, F. Corresponding to what will be explored in the subsection A,
 B and C,
 we have also four cases taken into consideration as follow.

Case B1, without classical knowledge of either state but with
 a single copy of  unknown states;

Case B2, with classical knowledge of one of the two states and
 a single copy of the other unknown state;

Case B3, with classical knowledge of one of the two states and the
absolute value of the inner product of both states, and also with a
single copy of the other unknown state;

Case B4, with classical knowledge of both states.

 The B1 and B4 cases will be investigated in subsection D and F,
respectively, and the B2 and B3 cases will be studied in subsection
E.
\subsection{Optimal unambiguous discrimination problems for Case A1}
In this subsection, we  first consider  the optimal unambiguous
discrimination problems for two qutrit states.  The preparing
probabilities is given, but none classical knowledge of
discriminated states is available.

The procedure of analysis is in line with Section II. A, but there
is some difference between them in how to construct the measurement
operators.

 Since we  have no classical knowledge
 about two qutrits $|\psi_{1}\rangle$ and $|\psi_{2}\rangle$,
 the right way of constructing POVM operators is  to  take advantage of the
 symmetrical
properties of the states. Denoting $|0\rangle$, $|1\rangle$ and
$|2\rangle$ as three vectors of a basis, we define the antisymmetric
states as follows
\begin{equation}
\label{3-A5-1a}
|\psi_{BC1}^{-}\rangle=\frac{1}{\sqrt{2}}(|0\rangle_{B}|1\rangle_{C}-|1\rangle_{B}|0\rangle_{C})
\end{equation}
and
\begin{equation}
\label{3-A5-1b}
|\psi_{BC2}^{-}\rangle=\frac{1}{\sqrt{2}}(|0\rangle_{B}|2\rangle_{C}-|2\rangle_{B}|0\rangle_{C})
\end{equation}
and
\begin{equation}
\label{3-A5-1c}
|\psi_{BC3}^{-}\rangle=\frac{1}{\sqrt{2}}(|1\rangle_{B}|2\rangle_{C}-|2\rangle_{B}|1\rangle_{C})
\end{equation}
and
\begin{equation}
\label{3-A5-2a}
|\psi_{AB1}^{-}\rangle=\frac{1}{\sqrt{2}}(|0\rangle_{A}|1\rangle_{B}-|1\rangle_{A}|0\rangle_{B})
\end{equation}
and
\begin{equation}
\label{3-A5-2b}
|\psi_{AB2}^{-}\rangle=\frac{1}{\sqrt{2}}(|0\rangle_{A}|2\rangle_{B}-|2\rangle_{A}|0\rangle_{B})
\end{equation}
and
\begin{equation}
\label{3-A5-2c}
|\psi_{AB2}^{-}\rangle=\frac{1}{\sqrt{2}}(|1\rangle_{A}|2\rangle_{B}-|2\rangle_{A}|1\rangle_{B})
\end{equation}
and introduce the projectors to the antisymmetric subspaces of the
corresponding qutrit as
\begin{equation}
\label{3-A6-1}
{P}_{BCi}^{as}=|\psi_{BCi}^{-}\rangle\langle\psi_{BCi}^{-}|
\end{equation}
and
\begin{equation}
\label{3-A6-2}
{P}_{ABi}^{as}=|\psi_{ABi}^{-}\rangle\langle\psi_{ABi}^{-}|
\end{equation}
with $i=1,2,3$. We now can take for ${\Pi}_{1}$ and ${\Pi}_{2}$
operators
\begin{equation}
\label{3-A7-1}
{\Pi}_{1}=\sum^{3}_{i=1}\lambda_{1i}I_{A}\otimes{P}_{BCi}^{as}
\end{equation}
and
\begin{equation}
\label{3-A7-2}
{\Pi}_{2}=\sum^{3}_{i=1}\lambda_{2i}{P}_{ABi}^{as}\otimes{I_{C}}
\end{equation}
To assure that ${\Pi}_{1}$, ${\Pi}_{2}$ and
${\Pi}_{0}=I-{\Pi}_{1}-{\Pi}_{2}$ be semi-positive operators, the
following constraints should be satisfied:
\begin{equation}
\label{3-A7} 2-\lambda_{1i}-\lambda_{2i}\geq0;\
1-\lambda_{1i}-\lambda_{2i}+\frac{3}{4}\lambda_{1i}\lambda_{2i}\geq0
\end{equation}
with $i=1,2,3$.

 Since
we have knowledge of $\eta_{1}$, our task is reduced to designing
$\lambda_{1i}(\eta_{1})$ and $\lambda_{2i}(\eta_{1})$ such that the
following average success probability
\begin{equation}
\label{3-A8} P=\eta_{1}p_{1}+(1-\eta_{1})p_{2}
\end{equation}
is maximal with the constrains given by Eq. (\ref{3-A7}).

  In this case, the corresponding optimal action parameters
are given by
\begin{equation}
\label{3-A-9a} \lambda_{1i}^{0,opt}(\eta_{1})=\
\bigg\{\begin{array}{cc}
0&\eta_{1}\leq{\frac{1}{5}}\\
\frac{2}{3}[2-\sqrt{\frac{1-\eta_{1}}{\eta_{1}}}]&\frac{1}{5}\leq{\eta_{1}}\leq{\frac{4}{5}}\\
1&\eta_{1}\geq{\frac{4}{5}}
\end{array}
\end{equation}
and
\begin{equation}
\label{3-A-9b} \lambda_{2i}^{0,opt}(\eta_{1})=\
\bigg\{\begin{array}{cc}
1&\eta_{1}\leq{\frac{1}{5}}\\
\frac{2}{3}[2-\sqrt{\frac{\eta_{1}}{1-\eta_{1}}}]&\frac{1}{5}\leq{\eta_{1}}\leq{\frac{4}{5}}\\
0&\eta_{1}\geq{\frac{4}{5}}
\end{array}
\end{equation}
with $i=1,2,3$ and the optimum success probability $P_{0}^{opt}$ can
be computed as follows
\begin{equation}
\label{3-A-9}
\sum^{3}_{i=1}\lambda_{1i}^{0,opt}\langle\Psi_{1}^{in}|I_{A}\otimes{P}_{BCi}^{as}|\Psi_{1}^{in}\rangle+
\sum^{3}_{i=1}\lambda_{2i}^{0,opt}\langle\Psi_{2}^{in}|{P}_{ABi}^{as}\otimes{I_{C}}|\Psi_{2}^{in}\rangle
\end{equation}
where the subscript $0$ of $P_{0}^{opt}$ means that we have no
\emph{a priori} classical knowledge of $|\psi_{1}\rangle$ and
$|\psi_{2}\rangle$.

\textbf{Remark:} From the aforementioned analysis, we reveal that
the measurement operators for discriminating two qutrit states are
more complicated than those for the qubit case. In general, it is
also impossible to express the optimum success probability
$P_{0}^{opt}$ as the function of $\beta$ and $\eta_{1}$.

\subsection{Optimal unambiguous discrimination problems for A2 and A3}
In this subsection, we studied the optimal unambiguous
discrimination problem of for the   {cases} A2 and A3.

The analysis is similar with  Section II. B, but the key challenge
is how to construct the measurement operators for qutrit states.

 Since we  know nothing
 about $|\psi_{2}\rangle$ but have the classical knowledge of $|\psi_{1}\rangle$,
 the right way of constructing POVM operators is  to  take advantage of the
 symmetrical
properties of the state as well as the classical knowledge of
$|\psi_{1}\rangle$. Denote $|\psi_{1}\rangle=|0\rangle$, and Let
$|0\rangle$, $|1\rangle$ and $|2\rangle$ constitute a basic basis.
We define the antisymmetric state
\begin{equation}
\label{3-5a}
|\psi_{AB1}^{-}\rangle=\frac{1}{\sqrt{2}}(|0\rangle_{A}|1\rangle_{B}-|1\rangle_{A}|0\rangle_{B})
\end{equation}
and
\begin{equation}
\label{3-5b}
|\psi_{AB2}^{-}\rangle=\frac{1}{\sqrt{2}}(|0\rangle_{A}|2\rangle_{B}-|2\rangle_{A}|0\rangle_{B})
\end{equation}
and introduce the projectors to the antisymmetric subspaces of the
corresponding qubit as
\begin{equation}
\label{3-6}
{P}_{ABi}^{as}=|\psi_{ABi}^{-}\rangle\langle\psi_{ABi}^{-}|
\end{equation}
with $i=1,2$.

By making full use of  the knowledge of $|\psi_{1}\rangle$,   we
construct the measurement operators ${\Pi}_{1}$ and ${\Pi}_{2}$ to
satisfy the no-error condition given by Eq.(\ref{A3}) as follows:
\begin{equation}
\label{3-7-1} {\Pi}_{1}=\sum^{2}_{i=1}\lambda_{1i}{P}_{ABi}^{as}
\end{equation}
and
\begin{equation}
\label{3-7-2}
{\Pi}_{2}=\sum^{2}_{i=1}\lambda_{2i}{|0\rangle_{A}|i\rangle_{B}{_{B}\langle}i|_{A}\langle{0}|}+\sum^{2}_{j,k=1}\lambda_{3jk}{|j\rangle_{A}|k\rangle_{B}{_{B}\langle}k|_{A}\langle{j}|}
\end{equation}
where  $\lambda_{1i}$, $\lambda_{2i}$ and $\lambda_{3jk}$ are
undetermined nonnegative real numbers. In terms of   Eqs.
(\ref{3-7-1}) and (\ref{3-7-2}), we can obtain the values of $p_{1}$
and $p_{2}$ by means of Eqs. (\ref{A2-1}) and (\ref{A2-2}).

By assuming that the  preparation probabilities of
$|\psi_{1}\rangle$ and $|\psi_{2}\rangle$ are $\eta_{1}$ and
$\eta_{2}$ (where $\eta_{2}=1-\eta_{1}$), respectively, we can still
define the average probability $P$ of successfully discriminating
two states as
\begin{equation}
\label{3-10} P=\eta_{1}p_{1}+(1-\eta_{1})p_{2}
\end{equation}
and our task is to maximize the performance Eq. (\ref{3-10}) subject
to the constraint that ${\Pi}_{0}=I-{\Pi}_{1}-{\Pi}_{2}$ is a
positive operator. After some calculations, we have
\begin{equation}
\label{3-10a}
p_{1}=\sum^{2}_{i=1}\lambda_{1i}|\langle\psi_{2}|i\rangle|^{2}
\end{equation}
and
\begin{equation}
\label{3-10b}
p_{2}=\sum^{2}_{i=1}\lambda_{2i}\beta^{2}|\langle\psi_{2}|i\rangle|^{2}+\sum^{2}_{j,k=1}\lambda_{3jk}|\langle\psi_{2}|j\rangle|^{2}|\langle\psi_{2}|k\rangle|^{2}
\end{equation}

To assure that operators ${\Pi}_{0}$, ${\Pi}_{1}$ and ${\Pi}_{2}$
are positive operators, we have the following inequality
constraints:

\begin{equation}
\label{3-i1}
1-{\lambda_{1i}}-{\lambda_{2i}}+\frac{1}{2}{\lambda_{1i}}{\lambda_{2i}}\geq0
\end{equation}
and
\begin{equation}
\label{3-i2}
0\leq{\lambda_{1i},\lambda_{2i},\lambda_{3jk}}\leq1(i,j,k=1,2)
\end{equation}

Based on the aforementioned observations,  we can give some further
analysis. Subsequently, we will discuss our strategies  for the A2
and A3 cases.

\textbf{(i)} For the A2 case,  we have the knowledge of preparing
probability $\eta_{1}$, but no knowledge of $\beta$.

Our strategy  is to design $\lambda^{1,w\beta}_{1i}(\eta)$,
$\lambda^{1,w\beta}_{2i}(\eta)$ and $\lambda^{1,w\beta}_{3jk}(\eta)$
to maximize the minimal performance
\begin{equation}
\label{3-H10-1}
J=\max\min_{\{\beta\}}[\eta_{1}p_{1}+(1-\eta_{1})p_{2}]
\end{equation}
subject to the  constraints described by Eqs. (\ref{3-i1})  and
(\ref{3-i2}).

 No matter what $\eta_{1}$ is, one should  always choose
$\lambda^{1,w\beta}_{3jk}(\eta_{1})=1$. Fortunately, this problem
can be reduced to maximizing the minimal performance
\begin{equation}
\label{3-H10-2}
J_{i}=\max\min_{\{\beta\}}[\frac{1}{2}\eta_{1}\lambda_{1i}+(1-\eta_{1})\lambda_{2i}\beta^{2}]
\end{equation}
subject to the  constraints described by Eqs. (\ref{3-i1}) and
\begin{equation}
\label{3-i3} 0\leq{\lambda_{1i},\lambda_{2i}}\leq1(i,j,k=1,2)
\end{equation}

 After some calculation,  we have
\begin{equation}
\label{3-24-a} \lambda_{1i}^{1,w\beta}(\eta_{1})=\
\bigg\{\begin{array}{cc}
0&\eta_{1}\leq{\frac{1}{2}}\\
2(1-\sqrt{\frac{1-\eta_{1}}{\eta_{1}}})&\frac{1}{2}\leq{\eta_{1}}\leq{\frac{4}{5}}\\
1&\eta_{1}\geq{\frac{4}{5}}
\end{array}
\end{equation}
and
\begin{equation}
\label{3-24-b} \lambda_{2i}^{1,w\beta}(\eta_{1})=\
\bigg\{\begin{array}{cc}
1&\eta_{1}\leq{\frac{1}{2}}\\
2-\sqrt{\frac{\eta_{1}}{1-\eta_{1}}}&\frac{1}{2}\leq{\eta_{1}}\leq{\frac{4}{5}}\\
0&\eta_{1}\geq{\frac{4}{5}}
\end{array}
\end{equation}
and
\begin{equation}
\label{3-24-c} \lambda^{1,w\beta}_{3jk}(\eta_{1})=1
\end{equation}

 By substituting
$\lambda^{1,w\beta}_{1i}(\eta_{1})$,
$\lambda^{{1,w\beta}}_{2i}(\eta_{1})$ and
$\lambda^{1,{w\beta}}_{3jk}(\eta_{1})$ into (\ref{3-10}), we obtain
the actual optimum success probability $P_{1}^{w\beta}$in this
strategy:
%\begin{equation}
%\label{3-24-1} P_{1}^{w\beta}(\beta,\eta_{1})=\
%\bigg\{\begin{array}{cc}
%P_{1_1}^{w\beta}(\beta,\eta_{1})&\eta_{1}\leq{\frac{1}{2}}\\
%P_{1_2}^{w\beta}(\beta,\eta_{1})&\frac{1}{2}\leq{\eta_{1}}\leq{\frac{4}{5}}\\
%P_{1_3}^{w\beta}(\beta,\eta_{1})&\eta_{1}\geq{\frac{4}{5}}
%\end{array}
%\end{equation}
%with
%\begin{equation}
%\label{24-1a}
%P_{1_1}^{w\beta}(\beta,\eta_{1})=(1-\eta_{1})(1-{\beta}^{2}))
%\end{equation}
%and
%\begin{equation}
%\label{24-1b}
%P_{1_2}^{w\beta}(\beta,\eta_{1})=[1+{\beta}^{2}(1-\eta_{1})-(1+{\beta}^{2})\sqrt{\eta_{1}(1-\eta_{1})}](1-{\beta}^{2}
%\end{equation}
%and
%\begin{equation}
%\label{24-1c} P_{1_3}^{w\beta}(\beta,\eta_{1})=
%(1-\frac{1}{2}\eta_{1}-{\beta}^{2}(1-\eta_{1}))(1-{\beta}^{2})
%\end{equation}
 where  the subscript $1$ of $P_{1}^{w\beta}$ means that we just
have \emph{a priori} classical knowledge of $|\psi_{1}\rangle$, one
of two discriminated states, and the superscript $w\beta$ of
$P_{1}^{w\beta}$ implies that the  optimum success probability is
obtained when making the decision based on the worst case for
$\beta$.

\textbf{Remark:} In this case, we can still choose the parameters of
the measurement operators based on the knowledge of \emph{a priori}
probability of the discriminated states. It should be pointed out
that the inner product of two discriminated qutrit states still
plays the the same role in optimum unambiguous state discrimination
 problems as that of two qubit states.

\textbf{(ii)} With \emph{a priori} classical knowledge of both
$|\langle\psi_{1}|\psi_{2}\rangle|=\beta$  and $\eta_{1}$ in hand,
our task in the third case is to get the optimum values
$\lambda^{1+,opt}_{1i}(\beta,\eta_{1})$,
$\lambda^{1+,opt}_{2i}(\beta,\eta_{1})$ and
$\lambda^{1+,opt}_{3jk}(\beta,\eta_{1})$ to optimize the  average
success probability
\begin{equation}
\label{3-A10} P=\eta_{1}p_{1}+(1-\eta_{1})p_{2}
\end{equation}
subject to the  constraints described by Eqs. (\ref{3-i1})  and
(\ref{3-i2}).

 No matter what $\eta_{1}$ is, one should  always choose
$\lambda^{1+,opt}_{3jk}(\eta_{1},\beta)=1$. Fortunately, this
problem can be reduced to choosing
$\lambda^{1+,opt}_{1i}(\beta,\eta_{1})$ and ,
$\lambda^{1+,opt}_{2i}(\beta,\eta_{1})$ to maximize performance
\begin{equation}
\label{3-B10-2}
J_{i}=\frac{1}{2}\eta_{1}\lambda_{1i}+(1-\eta_{1})\lambda_{2i}\beta^{2}
\end{equation}
subject to the  constraints described by Eqs. (\ref{3-i1}) and
(\ref{3-i3}).

After some calculations,  we have
\begin{equation}
\label{3-A28a} \lambda^{1+,opt}_{1i}(\beta,\eta_{1})=\
\bigg\{\begin{array}{cc}
0&\eta_{1}\leq\frac{{\beta}^{2}}{1+{\beta}^{2}}\\
2(1-\beta\sqrt{\frac{1-\eta_{1}}{\eta_{1}}})&\frac{{\beta}^{2}}{1+{\beta}^{2}}\leq{\eta_{1}}\leq{\frac{4{\beta}^{2}}{1+4{\beta}^{2}}}\\
1&\eta_{1}\geq{{\frac{4{\beta}^{2}}{1+4{\beta}^{2}}}}
\end{array}
\end{equation}
and
\begin{equation}
\label{3-A28b} \lambda^{1+,opt}_{2i}(\beta,\eta_{1})=\
\bigg\{\begin{array}{cc}
1&\eta_{1}\leq\frac{{\beta}^{2}}{1+{\beta}^{2}}\\
2-\frac{1}{\beta}\sqrt{\frac{\eta_{1}}{1-\eta_{1}}}&\frac{{\beta}^{2}}{1+{\beta}^{2}}\leq{\eta_{1}}\leq{\frac{4{\beta}^{2}}{1+4{\beta}^{2}}}\\
0&\eta_{1}\geq{{\frac{4{\beta}^{2}}{1+4{\beta}^{2}}}}
\end{array}
\end{equation}
and
\begin{equation}
\label{3-A28c} \lambda^{1+,opt}_{3jk}(\beta,\eta_{1})\equiv1
\end{equation}

Taking Eq. (\ref{3-A10}) into consideration, we obtain the
corresponding optimum success probabilities
$P_{1+}^{opt}(\beta,\eta_{1})$:
%\begin{equation}
%\label{3-A28-1} P_{1+}^{opt}(\beta,\eta_{1})=\
%\bigg\{\begin{array}{cc}
%P_{1+_1}^{opt}(\beta,\eta_{1})&\eta_{1}\leq\frac{{\beta}^{2}}{1+{\beta}^{2}}\\
%P_{1+_2}^{opt}(\beta,\eta_{1})&\frac{{\beta}^{2}}{1+{\beta}^{2}}\leq{\eta_{1}}\leq{\frac{4{\beta}^{2}}{1+4{\beta}^{2}}}\\
%P_{1+_3}^{opt}(\beta,\eta_{1})&\eta_{1}\geq{{\frac{4{\beta}^{2}}{1+4{\beta}^{2}}}}
%\end{array}
%\end{equation}
%with
%\begin{equation}
%\label{A28-1a} P_{1+_1}^{opt}(\beta,\eta_{1})=
%(1-\eta_{1})(1-\beta^{2})
%\end{equation}
%and
%\begin{equation}
%\label{A28-1b}
%P_{1+_2}^{opt}(\beta,\eta_{1})=[1+{\beta}^{2}(1-\eta_{1})-2\beta\sqrt{\eta_{1}(1-\eta_{1})}](1-\beta^{2})
%\end{equation}
%and
%\begin{equation}
%\label{A28-1c}
%P_{1+_3}^{opt}(\beta,\eta_{1})=[1-\frac{1}{2}\eta_{1}-\beta^{2}(1-\eta_{1})](1-\beta^{2})
%\end{equation}
 where the subscript $1+$ of $P_{1+}^{opt}$ means
that we  have \emph{a priori} classical knowledge of   one of the
two discriminated states and the absolute value of the inner product
of the two states.

\textbf{Remark:}  It is interesting to underline that it is
impossible to express the optimum success probability $P_{1+}^{opt}$
as the function of the inner product of two qutrit states $\beta$
and \emph{a priori} preparing probability $\eta_{1}$, but one can
make the optimum decision just based on the knowledge of $\beta$ and
$\eta_{1}$.

\subsection{Optimal unambiguous discrimination problems for Case A4}

If we have complete \emph{a priori} classical knowledge of both
$|\psi_{1}\rangle$ and $|\psi_{2}\rangle$, the measurement is
performed on  the detected qutrit.  One can select the detection
operators as follows:

(1) Select $|0\rangle=|\psi_{1}\rangle$, and choose another two
state $|1'\rangle$ and  $|2'\rangle$  so that the three states
$|0\rangle$, $|1'\rangle$ and  $|2'\rangle$ constitute a set of
basis base.

(2) Express  $|\psi_{2}\rangle$  in terms of $|0\rangle$,
$|1'\rangle$ and  $|2'\rangle$ as follows:
\begin{equation}
\label{3-b7}
|\psi_{2}\rangle=\cos\frac{\theta_{1}}{2}|0\rangle+e^{i\phi'_{1}}\sin\frac{\theta_{1}}{2}\cos\frac{\theta_{2}}{2}|1'\rangle+e^{i\phi'_{2}}\sin\frac{\theta_{1}}{2}\sin\frac{\theta_{2}}{2}|2'\rangle
\end{equation}
By setting
\begin{equation}
\label{3-b7a}
|1\rangle=e^{i\phi'_{1}}\cos\frac{\theta_{2}}{2}|1'\rangle+e^{i\phi'_{2}}\sin\frac{\theta_{2}}{2}|2'\rangle
\end{equation}
\begin{equation}
\label{3-b8}
|2\rangle=e^{i\phi'_{2}}\sin\frac{\theta_{2}}{2}|1'\rangle-e^{i\phi'_{1}}\sin\frac{\theta_{2}}{2}|2'\rangle
\end{equation}
we have
\begin{equation}
\label{3-b9}
|\psi_{2}\rangle=\cos\frac{\theta_{1}}{2}|0\rangle+\sin\frac{\theta_{1}}{2}|1\rangle
\end{equation}
and the three states $|0\rangle$, $|1\rangle$ and  $|2\rangle$
constitute another set of basis base.

(3)
\begin{equation}
\label{3-c1a} {\Pi}_{1}=\lambda_{1}{|e_{1}\rangle\langle{e_{1}}|}
\end{equation}
with
$|e_{1}\rangle=\sin\frac{\theta_{1}}{2}|0\rangle-\cos\frac{\theta_{1}}{2}|1\rangle$
and
\begin{equation}
\label{3-c1b} {\Pi}_{2}=\lambda_{2}{|1\rangle\langle1|}
\end{equation}

Denote $\cos\frac{\theta_{1}}{2}=\beta$, we have
$|\langle\psi_{1}|\psi_{2}\rangle|=\beta$. Our task is still to
choose $\lambda_{1}$ and $\lambda_{2}$ based on \emph{a priori}
information such that the average success probability given by Eq.
(\ref{c2}) is maximized.

To assure that ${\Pi}_{0}=I-{\Pi}_{1}-{\Pi}_{2}$, ${\Pi}_{1}$ and
${\Pi}_{2}$ are positive operators, we still have the   inequality
constraints given by Eqs.(\ref{c3}-\ref{c4}).

Since we have knowledge of preparing probability $\eta_{1}$ and
$\beta$, we will make the  decision given by
Eqs.(\ref{28a}-\ref{28b})

 Furthermore, we can obtain the optimum success probability Eq. (\ref{28-1}) with Eqs.(\ref{28-1a}-\ref{28-1c})
where  the subscript $2$ of $P_{2}^{opt}$ still means that we  have
the classical knowledge of both discriminated states.

\textbf{Remark:}   It is interesting to point out that the optimal
unambiguous discrimination problem for two qutrit states can be
reduce to the same one for two qubit states when  the classical
knowledge of both discriminated states is available.

\subsection{Optimal unambiguous discrimination problems for Case B1}
Since we have the same classical knowledge of discriminated states
in this case as in Section IV. A,  we can follow the analysis in
Section IV. A and choose ${\Pi}_{1}$ and ${\Pi}_{2}$ as Eqs.
(\ref{3-A7-1}) and (\ref{3-A7-2}).

To assure that ${\Pi}_{1}$, ${\Pi}_{2}$ and
${\Pi}_{0}=I-{\Pi}_{1}-{\Pi}_{2}$ be semi-positive operators, the
 constraints   on $\lambda_{1}$ and $\lambda_{2}$ described by Eq.(\ref{3-A7})
 should be satisfied.

However, since we have no knowledge of preparing probability, we
have to design $\lambda_{1}$ and $\lambda_{2}$ without \emph{a
priori} information of $\eta_{1}$. Our strategy is to maximize   the
minimal performance
\begin{equation}
\label{3-A-10}
J=P_{0}^{w\eta_{1}}=\max\min_{\{\eta_{1}\}}\{\eta_{1}p_{1}+(1-\eta_{1})p_{2}\}
\end{equation}
with the constraints in Eq. (\ref{3-A7}).

After careful calculations, we obtain that
\begin{equation}
\label{3-A-10a}
\lambda^{{0,w\eta_{1}}}_{1i}=\lambda^{0,w\eta_{1}}_{2i}=\frac{2}{3}
\end{equation}
Substituting Eq. (\ref{3-A-10a}) into Eq. (\ref{3-A-10}) yields
$P_{0}^{w\eta_{1}}$.
%\begin{equation}
%\label{A-11} P_{0}^{w\eta_{1}}
%\end{equation}

\subsection{Optimal unambiguous discrimination problems for Cases B2 and B3}
In this subsection, we will discuss the optimal unambiguous
discrimination problems for the B2  and B3 {cases} where partial
classical knowledge   but none  knowledge of preparing probabilities
of discriminated states are available.

Since we have the same partial classical knowledge of discriminated
states in this section as in Section IV.B,  we can follow the
analysis in Section IV.B and choose ${\Pi}_{1}$ and ${\Pi}_{2}$ as
Eqs.(\ref{3-7-1}) and (\ref{3-7-2}).

To assure that ${\Pi}_{1}$, ${\Pi}_{2}$ and
${\Pi}_{0}=I-{\Pi}_{1}-{\Pi}_{2}$ be semi-positive operators, the
 constraints on $\lambda_{1i}$,
$\lambda_{2i}$ and $\lambda_{3jk}$ described by (\ref{3-i1}) and
(\ref{3-i2}) should be satisfied.

Our task is to design$\lambda_{1i}$, $\lambda_{2i}$ and
$\lambda_{3jk}$ such that the average success probability
\begin{equation}
\label{B10} P=\eta_{1}p_{1}+(1-\eta_{1})p_{2}
\end{equation}
is maximized.

Subsequently, we will discuss our strategies for the B2 and B3
cases, respectively.

\textbf{(i)} If we  have neither the knowledge of preparing
probabilities nor the knowledge of $\beta$, our task is reduced to
designing $\lambda_{1i}^{1,w\beta\eta_{1}}$,
$\lambda_{2i}^{1,w\beta\eta_{1}}$ and
$\lambda_{3jk}^{1,w\beta\eta_{1}}$ to maximize the minimal
performance Eq.(\ref{B10}) subject to the  constraints in Eqs.
(\ref{3-i1})  and (\ref{3-i2}).

 No matter what $\eta_{1}$ is, one should  always choose
$\lambda^{1,w\beta\eta_{1}}_{3jk}(\eta_{1})=1$. Fortunately, this
problem can be further reduced to maximizing the minimal performance
\begin{equation}
\label{3-H10-2}
J_{i}=\max\min_{\{\beta,\eta_{1}\}}[\frac{1}{2}\eta_{1}\lambda_{1i}+(1-\eta_{1})\lambda_{2i}\beta^{2}]
\end{equation}
subject to the  constraints described by Eqs. (\ref{3-i1}) and
(\ref{3-i3}).

 Following some calculations in the subsection
IV. B, we have the optimal actions as follows
\begin{equation}
\label{3-BH10a} \lambda_{1i}^{1,w\beta\eta_{1}}=3-\sqrt{5}
\end{equation}
and
\begin{equation}
\label{3-BH10b}
\lambda_{2i}^{1,w\beta\eta_{1}}=\frac{1}{2}(3-\sqrt{5})
\end{equation}
and
\begin{equation}
\label{3-BH10c} \lambda_{3jk}^{1,w\beta\eta_{1}}=1
\end{equation}

 By substituting them  into Eq. (\ref{3-10}),
we get the  actual   optimum success probability
$P_{1}^{w\beta\eta_{1}}$ where the subscript $1$ of
$P_{1}^{w\beta\eta_{1}}$ means that we just have \emph{a priori}
classical knowledge of $|\psi_{1}\rangle$, one of two discriminated
states, and the superscript $w\beta\eta_{1}$ implies that the
optimum success probability is obtain based on the decision for  the
worst case for both $\beta$ and $\eta_{1}$.

\textbf{Remark:} Although the  actual   optimum success probability
$P_{1}^{w\beta\eta_{1}}$ depends on the both $\beta$ and $\eta_{1}$,
the optimum decision given by Eqs. (\ref{3-BH10a}-\ref{3-BH10c}) is
independent of  $\beta$ and $\eta_{1}$.

\textbf{(ii)} For the B3 case,  we have the knowledge of $\beta$,
but no knowledge of preparing probability $\eta_{1}$.

Our task is to design $\lambda^{1+,w\eta_{1}}_{1}(\beta)$,
$\lambda^{1+,w\eta_{1}}_{2}(\beta)$ and
$\lambda^{1+,w\eta_{1}}_{3}(\beta)$ to maximize the minimal
performance
\begin{equation}
\label{3-10-1}
J=\max\min_{\{\eta_{1}\}}[p_{1}\eta_{1}+p_{2}(1-\eta_{1})]
\end{equation}
subject to the  constraints given by Eqs. (\ref{i1})  and
(\ref{i2}).

 No matter what $\eta_{1}$ is, one should  always choose
$\lambda^{1,w\eta_{1}}_{3jk}(\beta)=1$. Fortunately, this problem
can be further reduced to maximizing the minimal performance
\begin{equation}
\label{3-H10-2}
J_{i}=\max\min_{\{\eta_{1}\}}[\frac{1}{2}\eta_{1}\lambda_{1i}+(1-\eta_{1})\lambda_{2i}\beta^{2}]
\end{equation}
subject to the  constraints described by Eqs. (\ref{3-i1}) and
(\ref{3-i3}).

Following  some similar calculations in subsection II. E, we have
\begin{equation}
\label{3-wA28a} \lambda_{1i}^{1+,w\eta_{1}}(\beta)=\
\bigg\{\begin{array}{cc}
1&\beta\leq\frac{\sqrt{2}}{2}\\
\beta^{2}+2-\sqrt{\beta^{4}+4\beta^{2}}&\beta\geq\frac{\sqrt{2}}{2}
\end{array}
\end{equation}
and
\begin{equation}
\label{3-wA28b} \lambda_{2i}^{1+,w\eta_{1}}(\beta)=\
\bigg\{\begin{array}{cc}
0&\beta\leq\frac{\sqrt{2}}{2}\\
\frac{3}{2}-\sqrt{\frac{1}{4}+\frac{1}{\beta^{2}}}&\beta\geq\frac{\sqrt{2}}{2}
\end{array}
\end{equation}
and
\begin{equation}
\label{3-wA28c} \lambda_{3jk}^{1+,w\eta_{1}}(\beta)\equiv1
\end{equation}

By substituting $\lambda^{1+,w\eta_{1}}_{1i}$,
$\lambda^{{1+,w\eta_{1}}}_{2i}$ and $\lambda^{1+,{w\eta_{1}}}_{3jk}$
into (\ref{3-10}), we obtain the actual success probability
$P_{1+}^{A}(\beta,\eta_{1})$ and optimum success probabilities in
the worst case $P_{1+}^{w\eta_{1}}$.

\textbf{Remark:} In this case, the optimum decision given by Eqs.
(\ref{3-wA28a}-\ref{3-wA28c}) is only the function of  $\beta$.

\subsection{Optimal unambiguous discrimination problems for case B4}

This subsection  discuss the optimal unambiguous discrimination
problem where complete classical knowledge of discriminated  states
but none   \emph{a priori} probabilities
 of preparing the discriminated states are available.

Here we have the same classical knowledge of discriminated states in
this case as in Section IV. C,  thus we can follow the analysis in
Section IV. C and choose ${\Pi}_{1}$ and ${\Pi}_{2}$ as Eqs.
(\ref{3-c1a}) and (\ref{3-c1b}).

In order to assure that ${\Pi}_{1}$, ${\Pi}_{2}$ and
${\Pi}_{0}=I-{\Pi}_{1}-{\Pi}_{2}$ be semi-positive, the
 constraints  on $\lambda_{1}$ and $\lambda_{2}$ given by Eqs. (\ref{c3}) and (\ref{c4}) should be satisfied
 where $|\langle\psi_{1}|\psi_{2}\rangle|=\beta$.

And what we shall do here is the same, i.e., to choose $\lambda_{1}$
and $\lambda_{2}$ based on \emph{a priori} information such that the
average success probability given by Eq.(\ref{c3}) is maximized with
the
 constraints in Eqs. (\ref{c3}) and (\ref{c4}).

When we have no knowledge of preparing probability $\eta_{1}$, our
task is to  choose $\lambda^{2,w\eta_{1}}_{1}(\beta)$ and
$\lambda^{2,w\eta_{1}}_{2}(\beta)$ to optimize the following
performance
\begin{equation}
\label{3-c2a}
J=\max\min_{\{\eta_{1}\}}[\lambda_{1}\eta_{1}+\lambda_{2}(1-\eta_{1})](1-\beta^{2})
\end{equation}
with the constraints described by Eqs. (\ref{c3}) and (\ref{c4}).

In this case, we still have
\begin{equation}
\label{3-W28}
\lambda^{2,w\eta_{1}}_{1}(\beta)=\lambda^{2,w\eta_{1}}_{2}(\beta)=\frac{1}{1+\beta}
\end{equation}
and
\begin{equation}
\label{3-W28-1} P_{2}^{w\eta_{1}}(\beta)=1-\beta
\end{equation}
where the subscript $2$ of $P_{2}^{w\eta_{1}}$ means that we  have
the classical knowledge of both discriminated states, and the
superscript $w\eta_{1}$ implies that the  optimum success
probability is defined in terms of the worst case for $\eta_{1}$.

\textbf{Remark:}   We would like to underline again that the optimal
unambiguous discrimination problem for two qutrit states can be
reduce to the same one for two qubit states when  the classical
knowledge of both discriminated states is available.

\section{Conclusion}
By comparing the results in Sect. IV with those in the Sect. II, we
would like to underline that the comprehensive analysis   for
unambiguously discriminating two qutrit states enhances the
principle viewpoint of the role of a priori information in the
optimum unambiguous state discrimination problems  in Sect. III.

Therefore, it has been  clarified in this paper that there are two
types of \emph{a priori} knowledge in optimum ambiguous state
discrimination problems: \emph{a priori} knowledge of discriminated
states themselves and \emph{a priori} probabilities of preparing
these states. It is demonstrated that both types of \emph{a priori}
knowledge can be utilized to improve the optimum average success
probabilities. It is very interesting to find that both types of
discriminators and the constraint conditions of action spaces are
decided just by the classical knowledge of discriminated states.
This is in contrast to the observation that both the loss functions
(optimum average success probabilities) and optimal decisions depend
on two types of \emph{a priori} knowledge.

It should be underlined that whether \emph{a priori} probabilities
of preparing discriminated states are available or not, what type of
discriminators one should design just depends on what kind of the
knowledge of discriminated states is provided. On the other hand,
how to choose the parameters of discriminators not only relies on
the \emph{a priori} knowledge of discriminated states, but also
depends on \emph{a priori} probabilities of preparing the states. In
conclusion, two kinds of \emph{a priori} knowledge can be utilized
to improve optimal performance but  play the different roles in the
optimization from the view point of decision theory.

When   considering the optimal unambiguous discrimination of
multiple linearly independent multiple-level quantum states, one
will have to realize that the complete classical knowledge of
discriminated states is almost the necessary condition for
constructing optimal unambiguous discriminator. This observation
further emphasizes the important role of \emph{a priori} classical
knowledge of the discriminated states in the optimal unambiguous
discrimination. In our opinion, the role of \emph{a priori}
knowledge in the optimization of quantum information processing
deserves further investigation.

\section{ACKNOWLEDGMENTS}
This work was funded by the National Natural Science Foundation of
China (Grant Nos. 60974037, 11074307). S. G. Schirmer acknowledges
funding from EPSRC. Z. Zhou and D. Hu acknowledge funding from
National Basic Research Program of China (2007CB311001). M. Zhang
and S. G. Schimer were also supported in part by the National
Science Foundation under Grant No. NSF PHY05-51164.

%\begin{biography}[{\includegraphics[width=1in,height=1.25in,clip,keepaspectratio]{mshell}}]{Michael Shell}

\end{document}